\documentclass[11pt,a4paper]{article}

\usepackage[margin=2.5cm]{geometry}
\usepackage{amsmath,amssymb}
\usepackage{graphicx}
\usepackage{booktabs}
\usepackage{array}
\usepackage{multirow}
\usepackage{caption}
\usepackage[colorlinks=true,linkcolor=blue,citecolor=blue,urlcolor=blue]{hyperref}
\usepackage{xcolor}
\usepackage[normalem]{ulem}
\usepackage{placeins}

\newcommand{\gam}{\ensuremath{\gamma}}
\newcommand{\gnbr}{\ensuremath{\gamma^{\mathrm{nbr}}}}

\graphicspath{{./}}

\title{A latent-space extrapolation grade built into graph atomic cluster expansion foundation potentials}

\author{%
Yury Lysogorskiy$^{1,*}$, Anton Bochkarev$^{1}$, Ralf Drautz$^{1}$\\[4pt]
\small $^{1}$ICAMS, Ruhr-Universit\"at Bochum, Bochum, Germany\\
\small $^{*}$Corresponding author: yury.lysogorskiy@rub.de}
\date{\today}

\hypersetup{allcolors=black}
\begin{document}
\maketitle

\begin{abstract}
Foundation machine-learning interatomic potentials cover broad configurational
and chemical spaces, but their reliability can vary across the atomic
environments encountered during a simulation.
Here we introduce the calibrated Mahalanobis (CALM) extrapolation grade \gam{}, a piecewise differentiable per-atom quantity integrated into GRACE foundation models and evaluated alongside energies and forces in a single model pass. 
We define $\gamma$ from nearest-cluster Mahalanobis distances in latent feature space, setting $\gamma=1$ from the training-distance distribution separately for each element
and cluster.
Controlled tests show that a normalized random projection of the invariant many-body basis detects structural and chemical extrapolation.
On different foundation datasets, OMat24 and SMAX, \gam{} correlates with atomic force errors and separates structures with different error distributions.
The CALM grade adds percent-level computational cost, and its spatial gradient guides uncertainty-biased data collection toward configurations with larger absolute force errors.
\end{abstract}

\section{Introduction}

Pretrained foundation machine-learning interatomic potentials (MLIPs) now
enable fast atomistic simulations across much of the periodic
table, with accuracy approaching their DFT references in tested domains.
Representative models include eSEN~\cite{fu2025esen},
GRACE~\cite{lysogorskiy2026graph,bochkarev2024grace},
MACE-MP~\cite{batatia2025foundation},
MatterSim~\cite{yang2024mattersim},
NequIP-OAM-XL~\mbox{\cite{tan2026nequip}},
ORB~\cite{orb},
PET-MAD~\cite{petmad2025}, and
SevenNet-Omni~\cite{kim2025sevennetomni}. Their training data draw on large
datasets including MPtrj~\cite{chgnet},
Alexandria~\cite{schmidt2023alexandria,wang2023alexandria2d},
OMat24~\cite{omat24}, MatPES~\cite{kaplan2025matpes},
MP-ALOE~\cite{kuner2025mpaloe}, MAD~\cite{mazitov2025mad},
SMAX~\cite{smax2026}, OMol25~\cite{levine2025omol25}, and
OPoly26~\cite{levine2026opoly26}.
This transferability is achieved by fitting flexible, often equivariant
representations to density-functional datasets that span a large part of
configurational and chemical space, so that a single parametrization can be
applied to systems for which it was never specifically fitted.
Fine-tuning can further improve accuracy for specific systems
with limited additional reference data~\cite{radova2025,liu2026tutorial,haenseroth2026}.
However, such models usually provide no native uncertainty quantification (UQ) method.
Thus, during a simulation there is no signal that distinguishes interpolation within the training distribution from silent extrapolation, where errors may grow by orders of magnitude. %

Extrapolation can be confined to defects, interfaces or reaction
events while the surrounding bulk remains within the training distribution.
Information-theoretic methods can identify novel individual atomic
environments~\cite{schwalbkoda2025}. For use at every step of molecular
dynamics, an extrapolation grade should likewise be atom-resolved, have a
calibrated threshold, and add little computational cost.

Deep ensembles are widely used for uncertainty estimation in MLIPs
and serve as an empirical reference method~\cite{tan2023}.
They estimate epistemic uncertainty from the disagreement between $N$
independently trained model replicas~\cite{lakshminarayanan2017}, so both
training and inference costs scale approximately with $N$.
This cost is substantial for foundation models whose individual training runs
consume thousands of GPU-hours~\cite{batatia2025foundation,medrano2026towards}.
Moreover, replicas trained on the same data and often using the same architecture
can make similar errors in unseen environments, so small ensemble disagreement
does not guarantee that a configuration lies within the training
distribution~\cite{kahle2022}.

Approximate Bayesian methods differ in their training and evaluation
requirements. Monte Carlo dropout, applied to interatomic potentials by Wen and
Tadmor~\cite{wen2020}, requires repeated stochastic forward passes. Stochastic
weight averaging Gaussian~\cite{maddox2019swag} collects weight statistics during
training, while improved variational online Newton~\cite{shen2024ivon} learns
an approximate weight distribution during optimization. The Laplace
approximation~\cite{daxberger2021laplace} can be applied after training,
but requires curvature information at the fitted parameters. Willow et
al.~\cite{willow2026bayesian} compared these latter three approaches with
ensembles for equivariant interatomic potentials.

Uncertainty estimates with lower evaluation cost are available from
resampled sparse Gaussian-process models~\cite{musil2019}, shallow ensembles
with shared network weights~\cite{kellner2024dpose}, and single-model evidential
regression~\cite{xu2026edl}. Resampling requires fitting models to subsets of the
training data, while the shallow-ensemble and evidential approaches incorporate
uncertainty estimation into the model architecture and training objective.
UQ methods for foundation MLIPs include readout ensembles and
quantile-regression readouts~\cite{bilbrey2025}, multi-head
committees~\cite{beck2025}, and ensembles of different pretrained foundation
models~\cite{liu2026}. The readout-based approaches require additional training
of the output layers. Heterogeneous ensembles reuse existing models without
additional training, but require force predictions from several foundation
models for each configuration.
The reliability of a supervised error predictor outside the domain covered
by its own labelled training data must be assessed separately.

Several foundation MLIPs now provide uncertainty estimates. PET-MAD uses
last-layer prediction rigidity (LLPR), which obtains post-hoc estimates
from a trained network's final features~\cite{bigi2024,chong2025}.
It calibrates the energy estimates against reference errors and samples
a last-layer ensemble to propagate uncertainty to derived
quantities~\cite{petmad2025}. Orb-v3 uses a confidence head to classify
per-atom force errors into 50 bins~\cite{orb}. Separately, POPS has been
applied to a linear corrector on top of a fixed MACE-MPA-0 foundation
model~\cite{swinburne2025pops,perez2025pops}.

Latent-space distances with cutoffs to control prediction error were studied
for neural-network chemistry models by Janet et al.~\cite{janet2019}.
Feature-based approaches for message-passing MLIPs
include per-atom GMM likelihoods on NequIP features~\cite{zhu2023} and
engineered invariant latent distances~\cite{musielewicz2024}.
Loss-trajectory analysis provides another single-model
approach~\cite{ltau2025}.
Several direct uncertainty estimates can be expressed as Mahalanobis-like
quadratic forms in method-specific features~\cite{grasselli2025}.
For a distance of the form
$(\mathbf z-\boldsymbol\mu)^{\mathsf T}\Sigma^{-1}
(\mathbf z-\boldsymbol\mu)$, weighting by the inverse covariance
$\Sigma^{-1}$ accounts for correlations and unequal variation along
feature directions.
The choice of features also matters: representations learned by different universal
potentials show substantial cross-model reconstruction errors~\cite{chorna2026latent}.

In this work we introduce the calibrated Mahalanobis (CALM) extrapolation grade
\gam{} for GRACE foundation potentials. We model the distribution of training
environments separately for each chemical element and calibrate the boundary
\gam{}$=1$ from the training distances. The features combine a random
projection of the normalized full invariant product basis with logarithmic
norm channels, which account for the wide range of basis magnitudes in the
foundation datasets.
We compare CALM with a deep ensemble and with the D-optimality (MaxVol)
grade~\cite{podryabinkin2017} evaluated on the same features. The latter was
previously compared with ensembles for atomic cluster expansion (ACE)
potentials~\cite{lysogorskiy2023}. Controlled tests show that the feature
choice determines whether the grade detects both structural and chemical
extrapolation. We then evaluate its relation to prediction error on OMat24
and SMAX datasets and use it to filter predictions on a materials-discovery benchmark.

\begin{figure}[!t]
\centering
\includegraphics[width=\textwidth]{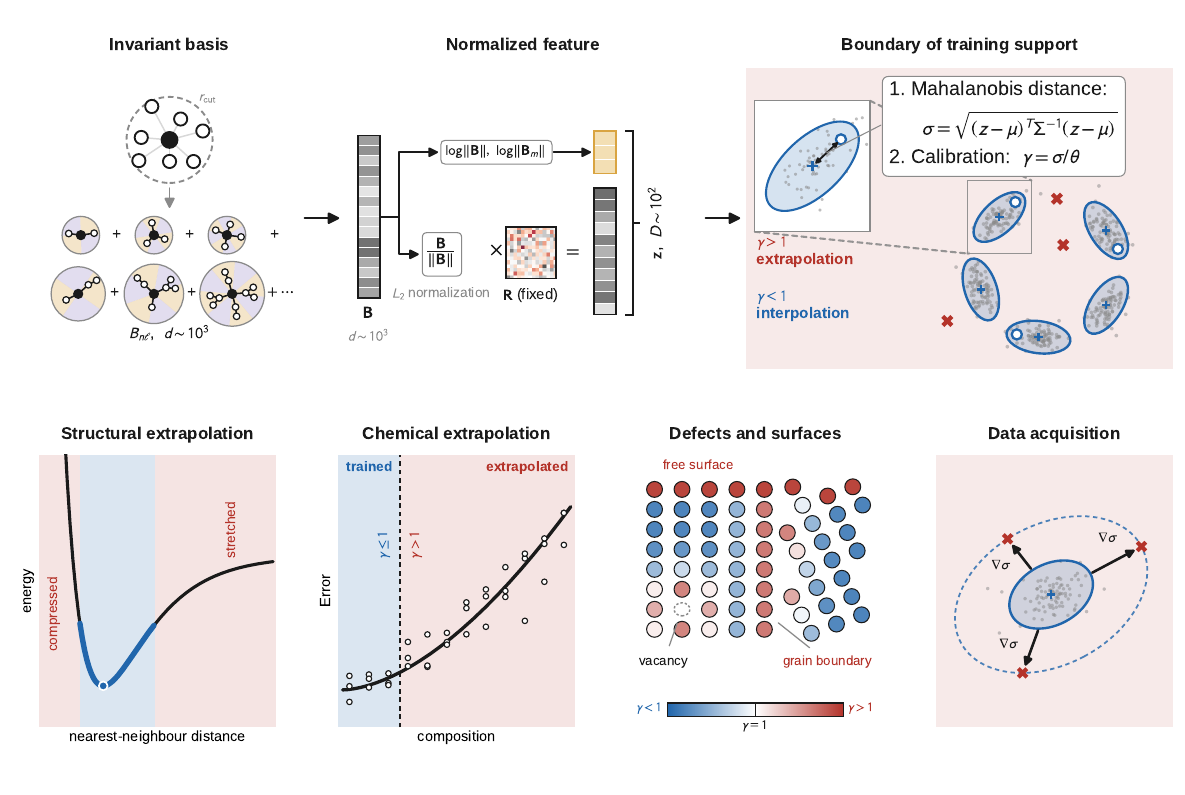}
\caption{\textbf{The extrapolation grade \gam{} and its applications.}
Schematic construction (top) and applications (bottom). The invariant basis is
normalized to unit $L_2$ norm and reduced by a fixed random projection. The
logarithmic norms of the full basis and of each reduction block are taken
before the normalization and appended to the projection. Per-element clustering
and nearest-cluster Mahalanobis scoring follow.
The calibrated value \gam{}$=1$ defines the
boundary of training support. Blue open circles lie inside this boundary and
red crosses outside it. The lower panels illustrate geometrical and chemical
extrapolation, atom-resolved defects and surfaces, and uncertainty-guided
sampling.}
\label{fig:overview}
\end{figure}

The CALM grade can be evaluated with energies only in a single model pass, with percent-level overhead in molecular dynamics (MD) in LAMMPS,
including its Kokkos GPU implementation. Within a fixed cluster assignment, the grade
is differentiable with respect to atomic positions. We use this gradient to
collect additional training configurations and compare the resulting potentials
with those fitted to the same number of structures from unbiased MD.
Fig.~\ref{fig:overview} summarizes the construction and selected applications.

\section{Results}

\subsection{Extrapolation grade from a calibrated Mahalanobis distance}

We construct CALM \textit{post-hoc} from a trained foundation MLIP and its training data
by fitting a clustered distribution of latent features separately for each
chemical element. Clustering has previously been used for stratified and
diversity-aware selection of MLIP training data~\cite{qi2024}.

For each chemical element $e$, an atom with latent feature vector $\mathbf z$ is
assigned to its nearest cluster centroid $k^{*}$ in feature space.
The Mahalanobis distance is first computed as
\begin{equation}
\sigma =
\sqrt{
(\mathbf z-\boldsymbol\mu_{e,k^{*}})^{\mathsf T}
\Sigma^{-1}_{e,k^{*}}
(\mathbf z-\boldsymbol\mu_{e,k^{*}})
},
\label{eq:mahalanobis}
\end{equation}
where $\boldsymbol\mu_{e,k^{*}}$ and $\Sigma_{e,k^{*}}$ are the mean and
covariance of the assigned cluster. The covariance accounts for the anisotropic
scatter of features within this cluster. To compare the resulting distances
across elements and clusters, we normalize them by a cluster-specific threshold:
\begin{equation}
\gam = \frac{\sigma}{\theta_{e,k^{*}}}.
\label{eq:gamma}
\end{equation}
The threshold $\theta_{e,k}$ is estimated from the
distribution of training-set distances $\sigma$ within each cluster as the
median plus three standard deviations estimated from the median absolute
deviation.
Across the pooled training atoms of the GRACE foundation models
considered here, 2--5\% lie above $\gam=1$.
Thus, $\gam=1$ defines the calibrated boundary of the training-data region for
the assigned cluster, with $\gam\leq1$ inside this region and $\gam>1$ outside it.
In this work, calibration denotes the placement of this threshold on the
training-set distance distribution, so that \gam{}$=1$ marks the upper tail of
the training distances by the same rule for every element and cluster. We
measure the relation between \gam{} and the prediction error separately,
through the correlations and conditional error distributions reported below.

Because the force on one atom depends on its neighbouring atomic environments,
we also examine the weighted neighbourhood-averaged extrapolation grade
\begin{equation}
    \gnbr_i
    =
    \frac{\gam_i+\sum_{j\in\mathcal N(i)}w_{ij}\gam_j}
         {1+\sum_{j\in\mathcal N(i)}w_{ij}},
    \qquad
    w_{ij}
    =
    \frac{1}{2}\left[1+\cos\!\left(\frac{\pi r_{ij}}{r_{\mathrm c,ij}}\right)\right].
    \label{eq:gamma-neighbour}
\end{equation}
Here $\mathcal N(i)$ is the model neighbour list and $w_{ij}$ is its cosine
cutoff weight. It was shown previously that spatial aggregation of atomic
uncertainty can improve the localization of force-error uncertainty~\cite{heid2024}.
We use the atomic grade \(\gam_i\) for calibration and sampling, and the
weighted neighbourhood-averaged grade \(\gnbr_i\) for the force-error evaluation.

The cluster means, covariance matrices, and $\sigma$ distributions are estimated
in three streaming passes over the training data. For large foundation datasets,
a representative subset is used for this purpose. The resulting distributions
and calibration thresholds are stored alongside the model. During
inference, \gam{} is evaluated from latent features $\mathbf z$ already generated
by the MLIP and requires only the cluster-distance calculation with the stored
inverse covariance matrices.
Within a fixed cluster assignment, \gam{} is differentiable with respect to the
atomic positions, and this gradient can define a biasing force for
uncertainty-driven sampling (see below).

\subsection{Minimal validation tests for an extrapolation grade}

We evaluate the extrapolation grade $\gam$ and its features $\mathbf z$ using
two tests. In the first, we vary the nearest-neighbour distance of few elemental
ground-state prototypes from compression through equilibrium to stretching.
The grade should identify the near-equilibrium region as interpolation and
both compression and stretching as extrapolation. In the second, we measure
how well the per-atom grade ranks atomic force errors on data within and
outside the training distribution.

To study the effect of different features and uncertainty models, we trained
three reference GRACE-2L models on the same element-stratified subset of
30{,}000 structures from OMat24, starting from independent random initializations
of the trainable weights. One model is used to construct and compare the
single-pass extrapolation grades, while all three form a deep ensemble reference.
We evaluate the force-error correlation on the held-out OMat24-val subset
(28{,}406 structures), which is drawn from the same parent dataset as
this training subset,
and on the out-of-distribution (OOD) SMAX subset (27{,}488
structures). These datasets test whether the extrapolation grade ranks force
errors near and beyond the OMat24 training domain.
Table~\ref{tab:ensemble} compares the response to compression and stretching
and the force-error correlations on OMat24-val and SMAX for all feature-space
and uncertainty-model combinations. It reports correlations for both the
atomic grade $\gam_i$ and the neighbourhood-averaged grade $\gnbr_i$.
The deep ensemble gives the highest correlations but requires three model evaluations. We use it as a reference for the single-pass estimators.

\begin{table}[t]
\centering
\footnotesize
\setlength{\tabcolsep}{6pt}
\caption{\textbf{Minimal validation tests of extrapolation grades.}
$D$ is the feature dimension. Geometrical extrapolation reports detection under
compression/stretching, using a threshold of $\gam=1$ or a
tenfold energy-spread increase for the ensemble. For OMat24-val
(in distribution) and SMAX (OOD), the columns give the Spearman correlations of
$\gam_i$ and $\gnbr_i$ with the atomic force error $\Delta F_i$. The bold feature
name marks the selected CALM construction.}
\label{tab:ensemble}
\begin{tabular}{@{}clcccccc@{}}
\toprule
 & & & \multicolumn{1}{c}{\shortstack{Test 1\\Geometrical\\extrapolation}}
 & \multicolumn{4}{c}{\shortstack{Test 2\\Force-error indication}} \\
\cmidrule(lr){4-4}\cmidrule(lr){5-8}
 & & & & \multicolumn{2}{c}{OMat24-val} & \multicolumn{2}{c}{SMAX} \\
\cmidrule(lr){5-6}\cmidrule(lr){7-8}
 & Feature space & $D$ & \shortstack{Compression/\\stretching}
  & $\rho(\gam_i,\Delta F_i)$ & $\rho(\gnbr_i,\Delta F_i)$
  & $\rho(\gam_i,\Delta F_i)$ & $\rho(\gnbr_i,\Delta F_i)$ \\
\midrule
\multirow{3}{*}{\rotatebox[origin=c]{90}{CALM}}
  & \textbf{Normalized feature} & 131 & yes/yes
  & 0.376 & 0.497
  & 0.539 & 0.617 \\
  & Linear projected basis & 128 & yes/no
  & 0.383 & 0.506
  & 0.600 & 0.654 \\
  & Hidden layer & 65$^{\dagger}$ & yes/yes
  & 0.184 & 0.327
  & 0.468 & 0.590 \\
\midrule
\multirow{3}{*}{\rotatebox[origin=c]{90}{D-opt}}
  & \shortstack[l]{Normalized, centred\\per cluster} & 131
  & yes/yes
  & 0.284 & 0.433
  & 0.514 & 0.607 \\
  & \shortstack[l]{Normalized, uncentred\\per element} & 131 & yes/yes
  & 0.178 & 0.295
  & 0.444 & 0.539 \\
  & \shortstack[l]{Hidden layer, uncentred\\per element} & 65$^{\dagger}$ & yes/yes
  & 0.077 & 0.140
  & 0.380 & 0.509 \\
\midrule
\multicolumn{2}{l}{Deep ensemble ($\times3$)} & --- & yes/yes
  & 0.722 & 0.688
  & 0.846 & 0.803 \\
\bottomrule
\addlinespace[2pt]
\multicolumn{8}{@{}p{0.98\linewidth}@{}}{\footnotesize$^{\dagger}$The hidden features are a deterministic function of the 17-channel readout bottleneck, so their effective dimension is at most 17.}
\end{tabular}
\end{table}

\subsection{Selection of feature space and uncertainty estimator}

The feature representation bounds the quality of a
feature-space extrapolation grade, since information lost in this
representation is unavailable to the estimator. We therefore test three
feature representations with the CALM model.

The first is the last hidden layer of the energy-readout multilayer perceptron
(MLP), which has $D=65$ features in the two-layer GRACE model architecture.
This representation is commonly used because the final
energy readout is linear in these features~\cite{zhu2023,bigi2024}.
However, in GRACE, the hidden features are a deterministic function of an
atomic-density bottleneck with $D_{\rho}=17$, which combines the 16 density
channels that enter the readout MLP with one channel that contributes to the
atomic energy linearly.
Their intrinsic dimension is therefore at most 17, while an environment with $N$ neighbours has $3N-3$ continuous geometric
degrees of freedom after global rotations are removed. For $N\geq7$,
geometrically distinct environments can in principle map to the same hidden
features. Such feature collapse (see the Supplementary Information for explicit
examples) may produce false negatives~\cite{vanamersfoort2021},
consistent with the comparatively weak correlations in Table~\ref{tab:ensemble}.

An alternative representation is the full invariant product basis $\mathbf B$,
obtained by concatenating the $l=0$ basis functions from all layers. For the
two-layer GRACE architecture used here, its dimension is
$D_{\mathrm{basis}}=7194$, which makes direct use impractical because
approximately $5\times10^{7}$ covariance entries would be required for every
element--cluster pair. We therefore map the basis
to $D=128$ using a fixed random projection that is drawn
once and approximately preserves pairwise distances between feature
vectors~\cite{johnson1984}, and term the resulting representation the
\emph{linear projected basis}. 
The basis parameterizes the manifold of local
atomic environments, whose intrinsic dimension lies far below
$D_{\mathrm{basis}}$, which motivates a random projection to far fewer
dimensions. The projection dimension is selected empirically below.

The linear projected basis gives the highest force-error correlations on both
evaluation sets (Table~\ref{tab:ensemble}). However, it fails the controlled
stretching test for all 6 elemental prototypes, as shown and
analysed in the Supplementary Information. As neighbours leave the cutoff, the
unnormalized basis and its linear projection approach the origin, which lies
inside their broad training distribution and may therefore be classified as in
distribution. In addition, the unnormalized basis norms span many orders of
magnitude and produce an excessively broad score range under compression,
reaching $1.0\times10^{18}$ on SMAX.

The third representation controls the scale of the basis by applying one
$L_2$ normalization before the random projection. The body-order blocks are
not normalized separately, so their relative magnitudes are preserved. In our
tests, this normalization improves the conditioning of the covariance matrices
by two to four orders of magnitude. The removed magnitude is included through
logarithmic norm channels for the full basis and its individual blocks. Together
with the 128 projected components, these channels give a 131-dimensional
\emph{normalized feature}. The norm channels also preserve the response as the
basis vanishes under stretching. The normalized feature detects both compression
and stretching, with a maximum grade of $9.9\times10^{3}$ across the two evaluation
sets. We therefore choose it despite the slightly better force-error ranking
of the linear projected basis.

The CALM model has two feature-space hyperparameters, the projection dimension
$D$ and the number of clusters per element $K$. The Supplementary Information
reports the force-error correlations and contamination at a fixed force-error
threshold for $D\in\{16,64,128,256\}$ and $K\in\{1,2,4,8,16\}$.
For this comparison we use the same $K$ for every element.
For the foundation models $K$ is being selected separately for each element using the elbow criterion described in the Methods.
Increasing $D$ and $K$ reduces the contamination of the interpolative
($\gam\leq1$) population, but also reduces the number of atoms assigned to it.
The Spearman correlation with force error peaks at four or eight clusters and
decreases at sixteen. Thus, the lower contamination at larger $K$ accompanies
a smaller interpolative population and does not indicate improved error ranking.
We select $D=128$ based on the force-error ranking on OMat24-val.

We test the sensitivity to the random projection using 3
independent seeds, re-clustering the training data and recalibrating the
thresholds for each projection. Across these draws, the Spearman correlation
with force error varies by at most 0.013, while the fraction of
atoms above \gam{}$=1$ varies by less than 1.3 percentage points.

Finally, we compare CALM with D-optimality on the same features and construct the
active sets either per element or per element-cluster pair, as described in the
Methods~\cite{podryabinkin2017,lysogorskiy2023}. For a square D-optimal active
set, the condition $\gamma_{\mathrm{Dopt}}\leq1$ defines a parallelotope in
feature space. The per-element D-optimality model on the hidden layer gives the
weakest correlation with force error in Table~\ref{tab:ensemble}. Using the
normalized features instead improves the correlations on both evaluation sets,
and cluster-centred active sets improve them further. The same comparison for a
last-layer prediction rigidity model~\cite{bigi2024,chong2025}, built on the
same hidden-layer feature, gives weaker correlations than either construction
and is reported in the Supplementary Information.

These improvements can be analysed geometrically in feature space. A single
cluster, for either CALM or D-optimality, defines one interpolation region that
may contain unsupported gaps between distinct modes of the training distribution.
For per-cluster D-optimality, the assigned cluster centroid is subtracted before
MaxVol selection and scoring, so that each cluster defines its own interpolative
parallelotope and cross-mode interpolation is reduced. This advantage of
resolving multiple modes is consistent with cluster- or class-dependent
treatments of heterogeneous spaces~\cite{morishita2023,ho2026} and with the
multi-active-set D-optimality scheme of Shuang et al.~\cite{shuang2026}.

Overall, the extrapolation grade from CALM correlates better with force error than
D-optimality constructed on the same normalized feature. Neighbourhood averaging
improves the force-error correlations of every single-pass extrapolation grade
without changing their ordering, whereas the ensemble correlations decrease
slightly (Table~\ref{tab:ensemble}). CALM is also more straightforward to parallelize
because the sufficient statistics used to construct the covariance matrices can
be accumulated by addition, whereas MaxVol requires coupled active-set selection.
We therefore choose CALM on the normalized invariant feature as the default
extrapolation grade for GRACE models.

\subsection{
Extrapolation threshold across datasets
}

\begin{figure}[!t]
\centering
\includegraphics[width=\textwidth]{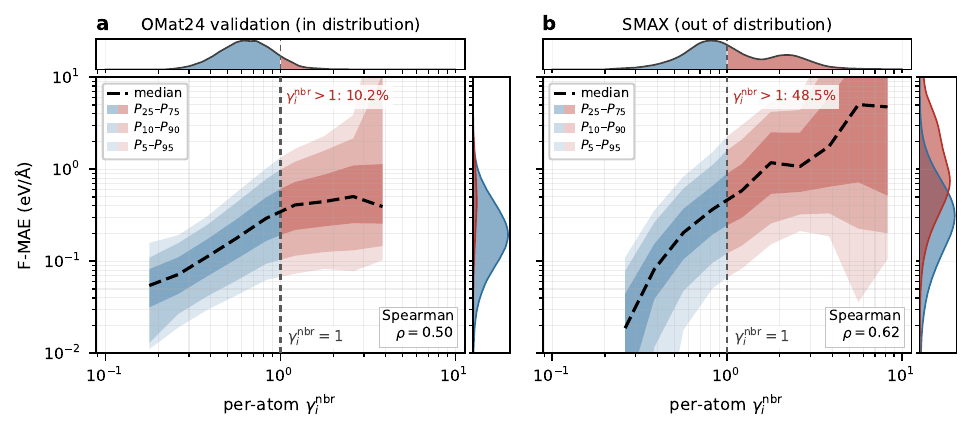}
\caption{\textbf{Reliability of the calibrated threshold.}
Per-atom force error against the neighbourhood-averaged grade \(\gnbr_i\) defined
by Eq.~(\ref{eq:gamma-neighbour}), for the controlled reference model on
\textbf{a}~held-out OMat24-val data and \textbf{b}~SMAX.
The black dashed line is the bin median, and the nested bands show the
$P_{25}$--$P_{75}$, $P_{10}$--$P_{90}$, and $P_{5}$--$P_{95}$ percentile
intervals. The upper strip is a smoothed density of the grade, and the right
strip the error density of each population separately.
The corresponding figure for the atomic grade \(\gam_i\) is given in the
Supplementary Information.}
\label{fig:reliability}
\end{figure}

Fig.~\ref{fig:reliability} compares the neighbourhood-averaged grade
$\gnbr_i$ with atomic force error on OMat24-val and SMAX, using the reference
GRACE model from the previous subsection. On both datasets, the median force
error increases with the grade over bins containing 99.96\% of
the OMat24-val atoms and 82.8\% of the SMAX atoms. The Spearman
correlation is 0.497 on OMat24-val and
0.617 on SMAX.
Most OMat24-val atoms fall in the two lowest grade bins, whereas SMAX atoms
span the full range, over which the median force error rises by more than an
order of magnitude (Supplementary Information). This broader range helps
explain the higher correlation on SMAX. At large grades, the median error
saturates while the upper tail continues to broaden. On both datasets, the
neighbourhood-averaged grade separates populations with different force-error
distributions at $\gnbr_i=1$. The corresponding analysis for the atomic grade
and the contamination curves are reported in the Supplementary Information.

\subsection{Structural and compositional extrapolation}

\begin{figure}[htbp]
\centering
\includegraphics[width=\textwidth]{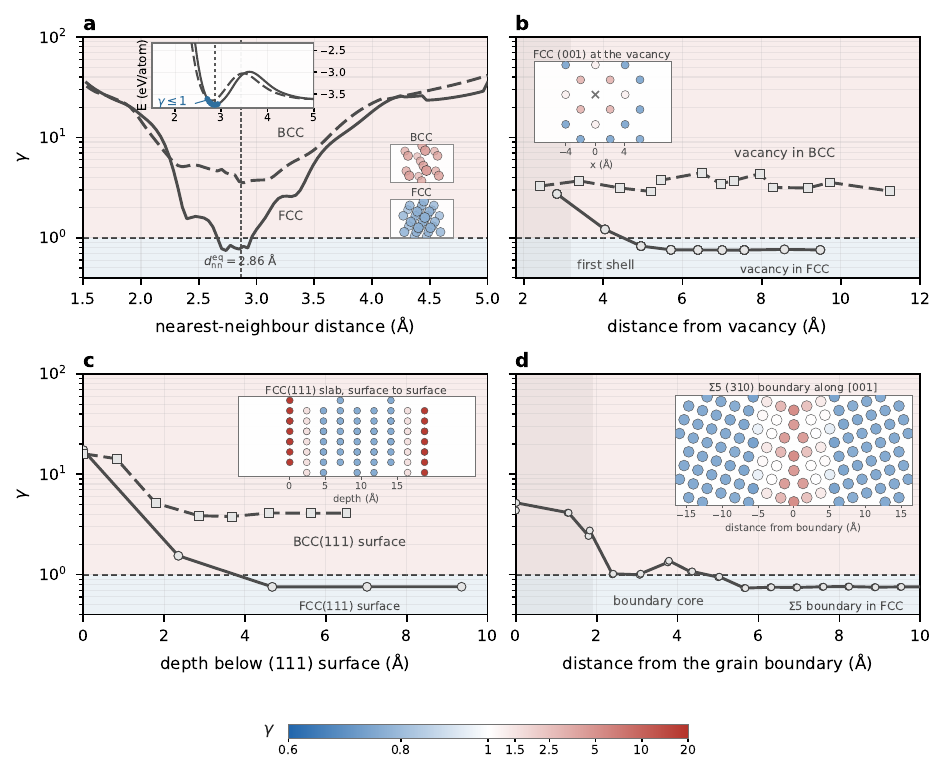}
\caption{\textbf{Geometrical extrapolation of an Al-FCC-only GRACE model.}
Solid curves denote
FCC and dashed curves BCC where both are shown.
\textbf{a}~Extrapolation grade along bond-length scans, with energy per atom inset,
\textbf{b}~Per-atom \gam{} around
a monovacancy. \textbf{c}~Layer-mean \gam{} below a free surface.
\textbf{d}~Per-atom \gam{} around a $\Sigma5$ (310) grain boundary. In the
structure insets and the two crystal icons, colour resolves \gam{} in space on the
scale below the figure.}
\label{fig:al-geom}
\end{figure}

We test structural extrapolation with a GRACE-1L model trained from randomly
initialized weights on near-equilibrium Al-FCC structures labelled by a
foundation GRACE model. We construct its CALM model from the same training
structures. Along the bond-length scans, $\gam\leq1$ only for FCC near its
equilibrium spacing, while the complete BCC scan has $\gam>1$
(Fig.~\ref{fig:al-geom}).
In the FCC vacancy and surface cells, the grade is largest near the missing
or truncated coordination shell and returns to its bulk value within a few
atomic layers. The grain boundary is also resolved locally. The BCC vacancy
and slab remain extrapolative throughout because BCC structures are absent
from the training set. Neighbourhood averaging gives the same spatial trends
with smoother variation between atoms. Further structural tests and comparisons
between the two grades are reported in the Supplementary Information.

We next consider compositional extrapolation using BCC Mo--W random solid
solutions. Mo and W are neighbouring BCC elements, and their equilibrium lattice
parameters differ by only 0.75\%, which largely removes geometrical
differences from this test. A GRACE model is trained on 90 BCC solid-solution
structures containing 128 atoms each and at most 25\% W, and evaluated over
$c_{\mathrm W}=0$--$100\%$.

\begin{figure}[!t]
\centering
\includegraphics[width=\textwidth]{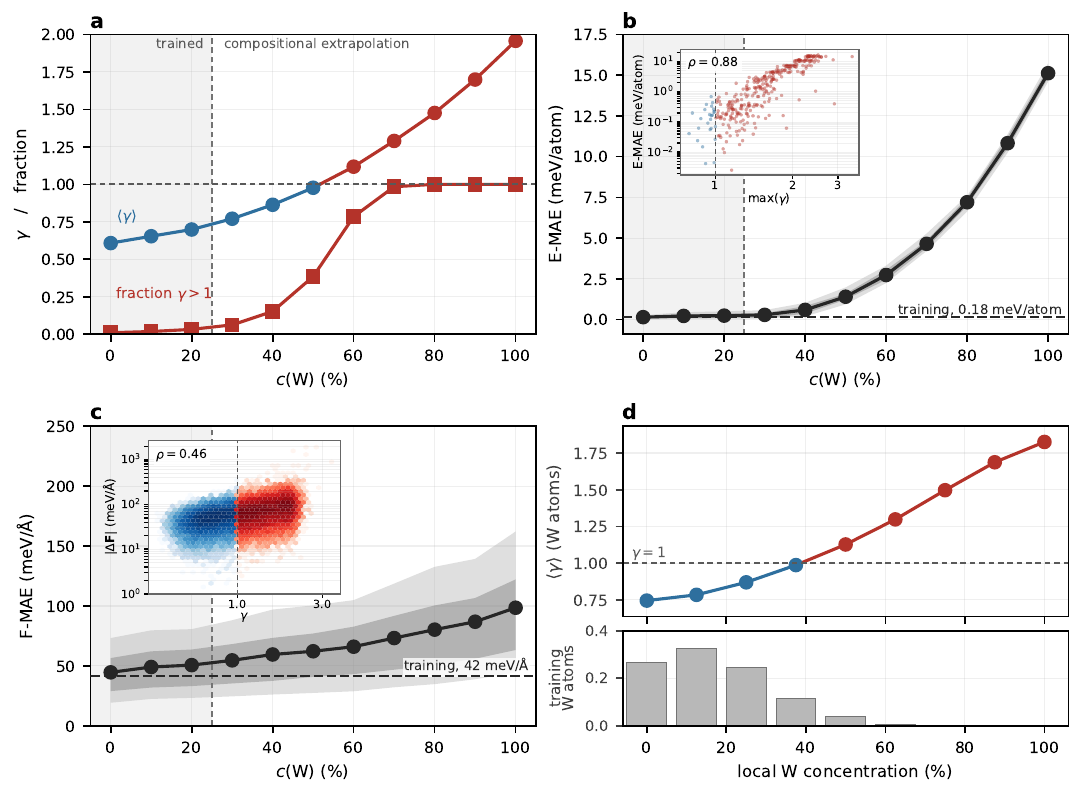}
\caption{\textbf{Chemical extrapolation in Mo--W solid solutions.}
The GRACE model was trained only on nominal concentration $c_{\mathrm W}\le25\%$, marked by the grey shaded
region. \textbf{a}~Mean \gam{} and the fraction of atoms with
$\gam>1$. \textbf{b}~Energy MAE per atom and \textbf{c}~force MAE, with the
25--75 and 10--90 percentile ranges and the error of the same model on its own training data
marked by the dashed horizontal line. \textbf{d}~Mean \gam{} of W atoms against the W
concentration of their first coordination shell, above the distribution of the
same quantity over the W atoms of the training data. Insets in \textbf{b} and \textbf{c} show the
corresponding structure- and atom-level correlations.}
\label{fig:chem}
\end{figure}

At the composition level, the mean \gam{} and flagged fraction increase towards
the W-rich limit (Fig.~\ref{fig:chem}a), together with the energy and force MAEs
(Fig.~\ref{fig:chem}b,c).
The mean grade of W atoms crosses one at a first-shell W concentration
of 37.5\% (Fig.~\ref{fig:chem}d). The nominal training limit of 25\% applies
to the whole cell and allows local shells with higher W concentrations.
The energy error rises above the training level only above a nominal W
concentration of 30\% (Fig.~\ref{fig:chem}b).
 The insets show that the correlation is preserved at
the level of individual structures and atoms, and does not arise only after
averaging over composition.

\subsection{Extrapolation in GRACE foundation models}
\label{sec:released-model-domains}

\begin{figure}[!t]
\centering
\includegraphics[width=\textwidth]{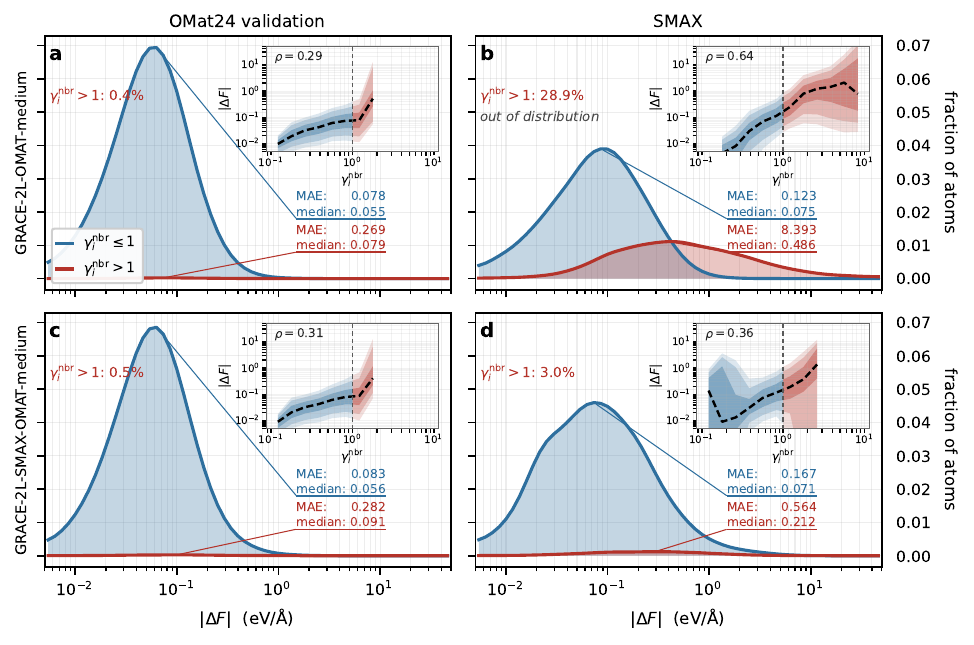}
\caption{\textbf{Training-data dependence of foundation-model uncertainty.}
Per-atom force-component errors are shown for atoms with $\gnbr_i\le1$ (blue) and $\gnbr_i>1$ (red). 
Rows compare GRACE-2L-OMAT-medium and GRACE-2L-SMAX-OMAT-medium;
columns compare OMat24-val and SMAX. 
Insets show error versus $\gnbr_i$, with medians (dashed), nested percentile bands, and per-atom Spearman correlation $\rho$.
Each panel reports the flagged fraction and both groups mean and median errors.
Averaging leaves few flagged atoms on OMat24-val.}
\label{fig:dfhist}
\end{figure}

We compare two foundation models of the same architecture trained on different
datasets: GRACE-2L-OMAT-medium~\cite{lysogorskiy2026graph} on OMat24 and GRACE-2L-SMAX-OMAT-medium~\cite{smax2026} on the
combined OMat24 and SMAX datasets (Fig.~\ref{fig:dfhist}). Both models assign
similarly low extrapolation rates to OMat24 validation data
(Fig.~\ref{fig:dfhist}a,c). On SMAX, the OMat24-only model flags
28.9\% of atoms, compared with
3.0\% for the model trained with SMAX
(Fig.~\ref{fig:dfhist}b,d).

For the OMat24-only model on SMAX, the force MAE increases from
0.123\,eV\,\AA$^{-1}$ below the calibrated boundary to
8.4\,eV\,\AA$^{-1}$ above it, where extreme OOD
configurations contribute a long error tail (Fig.~\ref{fig:dfhist}b).
The per-atom Spearman correlation is 0.638 on SMAX and
0.285 on OMat24-val (Fig.~\ref{fig:dfhist}a,b).
The lower extrapolation rate of the model trained with SMAX is consistent with
its broader training domain.

\subsection{Reliability filtering on a Materials Discovery benchmark}

We evaluate reliability filtering on the WBM formation-energy and
thermodynamic-stability benchmark~\cite{wang2021predicting} in Matbench
Discovery~\cite{riebesell2025framework}. WBM lies close to the training domain
of the benchmark model and tests error ranking in a predominantly interpolative
dataset. We use the maximum atomic $\gam$ as the structure-level grade, since
a single extrapolative atomic environment can affect the predicted energy of
the whole cell.

\begin{figure}[!t]
\centering
\includegraphics[width=\textwidth]{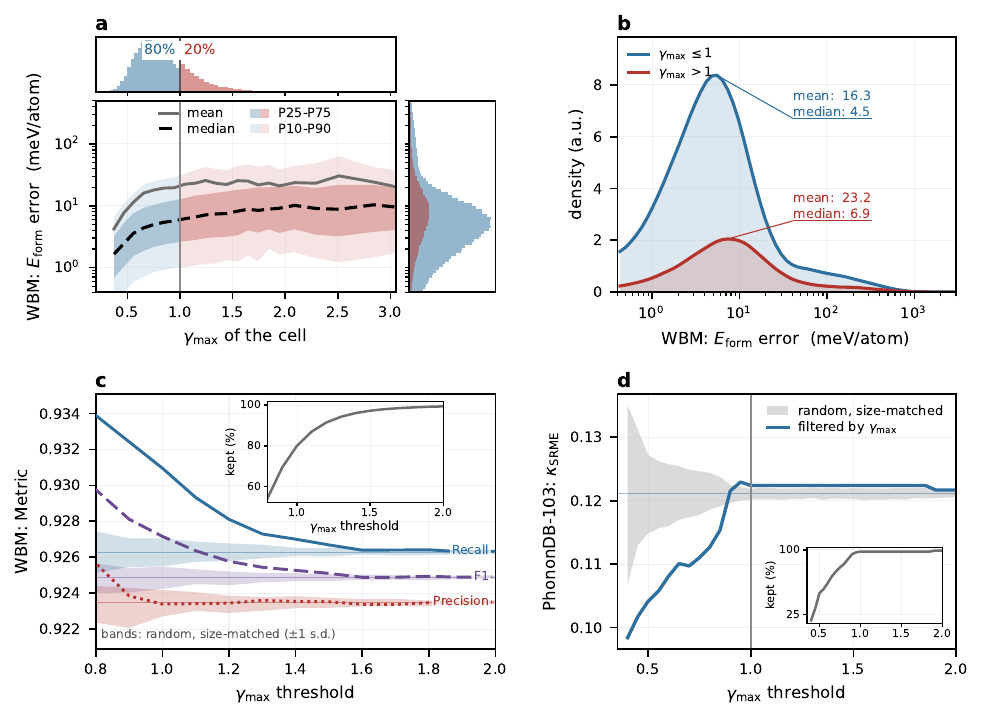}
\caption{\textbf{Extrapolation grade as a reliability criterion on a
Materials Discovery benchmark.} Extrapolation grade \gam{} of
GRACE-3L-OAM-L on the unique-prototype subset of the WBM test set. \textbf{a} Absolute formation-energy error versus the maximum atomic \gam{} per structure. Dashed black and grey lines show binned medians and means; shaded bands show $P_{25}$--$P_{75}$ and $P_{10}$--$P_{90}$ percentiles, blue for \gam{}$\le1$ and red for \gam{}$>1$. Marginal distributions are split at the same boundary. \textbf{b}~Error distributions split at \gam{}$=1$. 
\textbf{c}Discovery metrics versus the \gam{} threshold; bands show one standard deviation of size-matched random controls, the inset shows the retained fraction, and thin lines mark unfiltered values.
\textbf{d}~Thermal-conductivity error for interpolative PhononDB materials,
with the same size-matched random control.}
\label{fig:wbm}
\end{figure}

Fig.~\ref{fig:wbm}a shows the absolute formation-energy error of
GRACE-3L-OAM-L against the largest per-atom extrapolation grade of
each cell. The median error increases with \gam{} up to \(\gam=1\) and then
saturates. The score distribution is narrow, with 20.1\% of the cells outside this boundary.
Trajectories containing structures with matching WBM prototype labels
were excluded when the subsampled Alexandria (sAlex) training dataset
was constructed~\cite{omat24}.
Structures with $\gam>1$ have larger mean and median formation-energy errors
than those with $\gam\leq1$ (Fig.~\ref{fig:wbm}b). Filtering at $\gam=1$
reduces the formation-energy MAE from 17.7 to
16.3\,meV\,atom$^{-1}$ and keeps 80\% of the
structures (Fig.~\ref{fig:wbm}c).

Lower thresholds exclude more structures, so we compare the discovery metrics
of each filtered set with those of randomly sampled subsets of the same size
(Fig.~\ref{fig:wbm}c). Recall and F1 exceed the random controls by more than
one standard deviation, whereas precision remains within the random-control
interval. 

The benchmark also includes 103 PhononDB structures evaluated using
the material-averaged symmetric relative mean error of thermal conductivity,
$\kappa_\mathrm{SRME}$~\cite{pota2026thermal}. At $\gam=1$, filtering removes
only two materials, both predicted accurately, and slightly increases
$\kappa_\mathrm{SRME}$ (Fig.~\ref{fig:wbm}d). Lowering the threshold to
$\gam\leq0.5$ reduces the error from 0.1211 to
0.1041 while keeping 50\% of the materials.
The size-matched random control remains close to the unfiltered value (Supplementary Information).

\subsection{Uncertainty-guided data acquisition}

Uncertainty estimates can also guide the expansion of MLIP training datasets.
Schwalbe-Koda et al.~\cite{schwalbekoda2021} sampled new geometries along
the gradient of a differentiable ensemble uncertainty, and uncertainty-driven
dynamics accelerates active learning~\cite{kulichenko2023}.
Van der Oord et al.~\cite{vanderoord2023} introduced hyperactive learning
(HAL), in which the uncertainty gradient biases MD towards uncertain regions.
In active exploration with ACE potentials, the extrapolation grade was instead
maximized directly~\cite{lysogorskiy2023}. Zaverkin et
al.~\cite{zaverkin2024} further showed that uncertainty-biased MD can explore
extrapolative regions and rare events. Here we test whether sampling along the gradient of the GRACE extrapolation
grade improves a potential fitted to the collected structures compared with
unbiased MD sampling. In both cases, we upfit the potential by continuing
training from its existing weights on the extended dataset.

We perform a numerical experiment on aluminium data, in which 
GRACE-2L-SMAX-OMAT-large serves only as a replacement for DFT energies, forces
and stresses. The initial common seed dataset contains 50 configurations of a
108-atom FCC cell sampled by NPT-MD at 300 and 500\,K. A single-element GRACE-1L
potential and its CALM model are fitted to this dataset. Then, for reference,
unbiased NPT-MD is continued and 200 additional structures are selected. For
comparison, HAL at 300\,K biases the forces and cell along the gradient of
\gam{} and selects the same number of structures when $\gam_{\max}>1.5$, as
described in the Methods. In order to avoid repeated
sampling from visited regions of feature space, the CALM covariances are updated
after every collection so that \gam{} decreases locally and the walk can proceed
towards new configurations. The GRACE parameters remain fixed throughout
sampling. After collection, the initial potential is upfitted separately with
the MD or HAL structures and evaluated on common test
configurations.

\begin{figure}[!t]
\centering
\includegraphics[width=\textwidth]{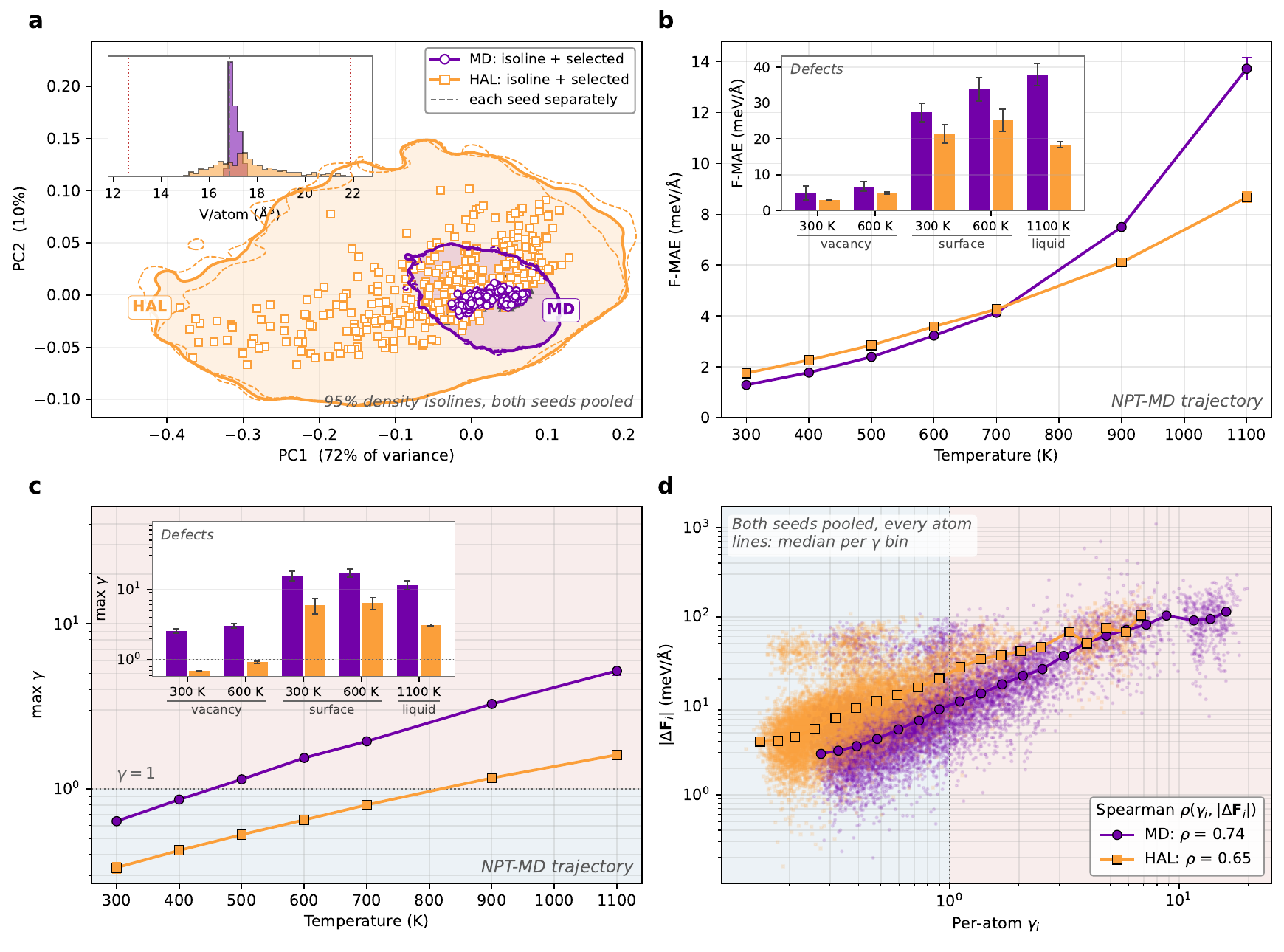}
\caption{\textbf{Uncertainty-guided data acquisition at a fixed labelling budget in aluminium.}
Purple and orange colors show MD and HAL sampling strategy correspondingly. \textbf{a} Collections projected onto the first two principal components of the seed potential’s features. Solid contours enclose 95\% of pooled atoms; dashed contours show each replicate. Symbols mark configuration centroids, and the inset shows volume per atom. 
\textbf{b} Force MAE on independent reference trajectories and, in the inset, held-out vacancy, $(111)$-surface, and liquid configurations. 
\textbf{c} Maximum atomic grade on the same configurations, using each potential’s own CALM model.
\textbf{d}~Atomic force error versus \gam{} across all evaluation sets.}
\label{fig:budget}
\end{figure}

To test reproducibility, we repeat the complete experiment with an independently
generated seed dataset and new random seeds throughout the workflow.
Fig.~\ref{fig:budget} combines the two replicates.

Fig.~\ref{fig:budget}a shows that the two strategies generate different
training distributions. The configurations drawn from unbiased MD remain in a
compact thermal region of the feature space and near the equilibrium volume.
In contrast, uncertainty-guided sampling covers a broader feature-space region
and much of the allowed volume interval, with most of the additional coverage
on the stretched side. The replicate-specific isolines cover nearly the same
region.

Fig.~\ref{fig:budget}b compares the two upfitted potentials on independent MD
at increasing temperatures and on held-out vacancy, $(111)$-surface and liquid
configurations. At lower temperatures, the potential fitted to the unbiased
collection is slightly more accurate below about 700\,K due to denser sampling
of this region in its training set. The ordering reverses at higher
temperatures, and at 1100\,K the force error decreases from 13.7 to
8.7\,meV\,\AA$^{-1}$ for the potential fitted to the uncertainty-guided
collection. On the liquid configurations, the error decreases from 37.9 to
18.4\,meV\,\AA$^{-1}$. The surface configurations show the same ordering,
whereas the relaxed vacancy remains close to thermal bulk and is described
similarly by both potentials. For the vacancy, however, the difference between
replicates is comparable to the difference between sampling strategies. It
is smaller for the other tests.

The CALM models generated for both potentials reflect the difference in their
covered training domains (Fig.~\ref{fig:budget}c). In both replicates, the
potential fitted to unbiased MD crosses the calibrated boundary between 400 and
500\,K, whereas the uncertainty-guided potential crosses it between 700 and
900\,K. Fig.~\ref{fig:budget}d further shows that the
atomic grades remain positively correlated with the corresponding force errors
after upfitting, with Spearman correlations of 0.74 and 0.65 for the MD and HAL
potentials, respectively.

For silicon, uncertainty-guided sampling moves towards compression in both
replicates, in contrast to the expansion observed in aluminium, consistent
with their different liquid-to-solid density ratios (Supplementary Information).
For aluminium, the additional structures improve the fitted potential on the
high-temperature and liquid tests at the same labelling cost, while unbiased
MD gives lower errors below about 700\,K. The benefit therefore depends on the
configurations collected and the conditions under which the potential is tested.

\subsection{Uncertainty-guided generation of candidate structures}

We next apply the HAL methodology to expand the training dataset of
GRACE-2L-SMAX-large, starting from 40 chemically diverse SMAX training
structures. Each seed is propagated at 1000\,K by thermostatted MD with an
additional bias force derived from the gradient of the extrapolation grade.
Because \gam{} also increases at short bond lengths, the cell volume is ramped
linearly from $V_0$ to $1.4V_0$ to bias the search towards tensile
configurations. 
The short-range guard, the collection criterion and the update of the
CALM covariances are given in the Methods.

For the unbiased control, the same seeds are propagated without the
uncertainty force and without CALM updates. DFT single-point calculations
following the SMAX settings completed for 78 of the 94 uncertainty-biased
configurations of the carried-mixture protocol and for 69 of the 76
corresponding control configurations.

\begin{figure}[!t]
\centering
\includegraphics[width=\textwidth]{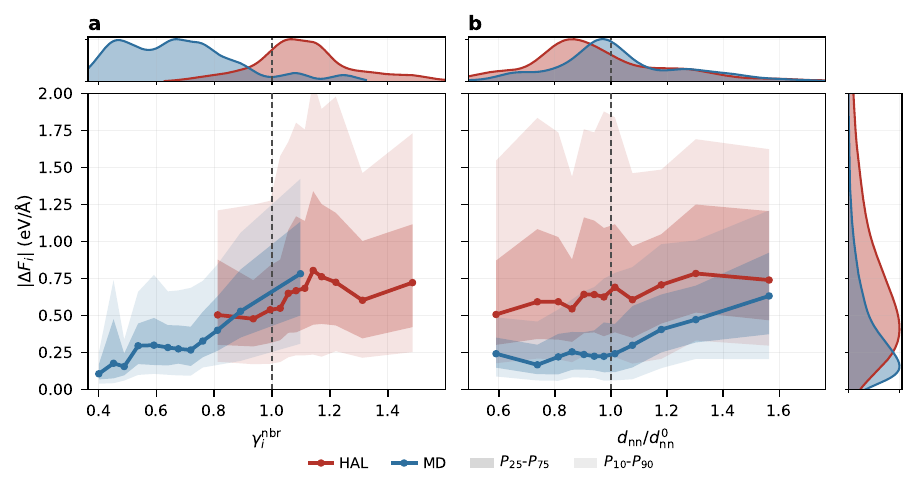}
\caption{\textbf{Collection of new training structures for foundation model: uncertainty-biased vs. thremal-driven control.} 
Configurations collected with GRACE-2L-SMAX-large by uncertainty bias (red) or a thermal walk (blue) used the same seeds and volume ramp and were labelled by spin-polarized DFT.
\textbf{a} Atomic force error versus \(\gnbr_i\), with medians and nested $P_{25}$--$P_{75}$ and $P_{10}$--$P_{90}$ percentile bands.
\textbf{b} The same error versus nearest-neighbour distance relative to the relaxed seed. 
Marginal densities show both horizontal variables and the error; dashed lines mark \(\gnbr_i=1\) and the relaxed spacing.
}
\label{fig:selfexp}
\end{figure}

Fig.~\ref{fig:selfexp} compares the per-atom force
error with \(\gnbr_i\) and the relative nearest-neighbour distance.
Across both collections, the extrapolation grade correlates positively with the
force error, with a Spearman correlation of $\rho=0.52$
(Fig.~\ref{fig:selfexp}a). HAL shifts the grade distribution upwards, from a
median \(\gnbr_i\) of 0.66 for MD to 1.10, and increases the mean per-atom force
error from 0.39 to 0.83\,eV\,\AA$^{-1}$. Despite these different distributions,
both collections follow the same error--\(\gnbr_i\) trend.

Fig.~\ref{fig:selfexp}b shows the correlation of force error with relative
nearest-neighbour distance. For unbiased MD, the median error increases from
0.23\,eV\,\AA$^{-1}$ near the relaxed spacing to
0.50\,eV\,\AA$^{-1}$ for strongly stretched
environments with $d_\mathrm{nn}/d_\mathrm{nn}^{0}>1.15$.
The HAL collection has larger errors throughout the sampled bond-length range,
reaching a median of 0.73\,eV\,\AA$^{-1}$ in the
strongly stretched region.
The HAL configurations satisfy the short-range guard and have larger grades
and force errors than the unbiased collection. They were collected without
retraining GRACE and provide candidates for extending its training dataset.

\subsection{Overhead in molecular dynamics}

The CALM extrapolation grade $\gam$ uses the same basis expansion as the energy and is evaluated
alongside energies and forces in a single model pass. Its measured overhead
ranges from about 6\% for the one-layer model at the smaller system size to
below 1\% for the three-layer model. The relative overhead decreases as the
cost of the underlying potential increases with model depth. Computing the gradients of $\sigma$ with respect to atomic
positions raises the overhead to about 30\% for the one- and two-layer
models. We timed the Kokkos and TensorFlow backends at two system sizes
(Supplementary Information).

\section{Discussion}
\label{sec:discussion}

The feature comparison shows that force-error correlation alone is
insufficient to select an extrapolation grade. Although the linear projected
basis gives stronger correlations than the normalized representation, it fails
to identify stretching as neighbours leave the cutoff. Normalization and the
logarithmic norm channels preserve the response to both compression and
stretching, at the cost of a modest reduction in error ranking.

Once the features are fixed, the differences between CALM and
D-optimality are smaller than those between feature representations. Both can
be related to Mahalanobis-type distances~\cite{grasselli2025}, and both benefit
from resolving distinct modes of the training distribution into separate
clusters. The improvement from centred, per-cluster D-optimality active sets
supports this interpretation. However, each CALM ellipsoid remains an
approximation to the occupied region of feature space and can include
environments absent from training. For elements with few training environments,
cluster merging and threshold fallback further limit the resolution of this
description.

Calibration gives $\gam=1$ a common interpretation across elements
and clusters by applying the same rule to their training-distance distributions,
with a small fraction of training environments above the boundary. The Al and
Mo--W tests show how the grade responds to structures and compositions absent
from the training data, while the foundation-model evaluations show that it
correlates with prediction error. This correlation can vary with dataset and
chemistry, as for other target-independent feature-space UQ
methods~\cite{grasselli2025}. Estimating error quantiles or the probability that
an error exceeds a chosen threshold for a given range of $\gam$ therefore
requires independent reference data relevant to the application domain.

The standard deviation of force predictions across the deep ensemble
correlates more strongly with force error (Table~\ref{tab:ensemble}), but comes
at the cost of parameterizing and evaluating multiple models. In contrast,
CALM uses features already computed for the energy and forces, so the grade
can be evaluated at every MD step with percent-level overhead.

The spatial gradient of $\gam$ also provides a bias for collecting
training configurations. Updating the CALM covariance matrices on the fly to
cover configurations already visited directs sampling towards new
configurations while the potential remains fixed. Training on these actively
selected configurations improves the potential's transferability to a broader
range of atomic environments, as demonstrated by the high-temperature and
liquid tests. A similar procedure applied to the foundation model selects
configurations with larger force errors than the unbiased control, as
confirmed by DFT calculations, and provides additional reference data for
extending its training dataset.

Overall, an uncertainty indicator should be part of every interatomic
potential to assess its reliability and transferability during simulations.
The results presented here show that such an indicator can be evaluated
together with energies and forces at little additional computational cost.

\section{Methods}

\subsection{Latent features and uncertainty models}

The latent representation used for CALM is constructed from the concatenated
$\ell=0$ invariant product basis along the atomic-energy path. For the
two-layer models, this basis has $D_{\mathrm{basis}}=7194$ components (3738
from the first interaction layer and 3456 from the second). We write the
concatenation for atom $i$ as
$\mathbf B_i=(\mathbf B_{i,1},\dots,\mathbf B_{i,n_{\mathrm{layers}}})$, where
$\mathbf B_{i,m}$ collects the components contributed by reduction layer $m$.
The feature map is then applied in three steps.

First, the basis is normalized to unit Euclidean norm,
\begin{equation}
    \hat{\mathbf B}_i
    =
    \frac{\mathbf B_i}{\lVert\mathbf B_i\rVert+\varepsilon_0},
    \label{eq:normalization}
\end{equation}
with $\varepsilon_0=10^{-12}$. To preserve the magnitude removed by
normalization, we record the logarithmic norms of the raw full basis and each
reduction block,
\begin{equation}
    d_{i,0}=\log\left(\lVert\mathbf B_i\rVert+\varepsilon_0\right),
    \qquad
    d_{i,m}=\log\left(\lVert\mathbf B_{i,m}\rVert+\varepsilon_0\right).
    \label{eq:density-channels}
\end{equation}
Here $d_{i,0}$ denotes the channel of the full basis and $d_{i,m}$ that of
reduction block $m$, with $m=1,\dots,n_{\mathrm{blocks}}$ and one reduction
block per interaction layer.

Second, the normalized basis is projected onto $D_{\mathrm{rp}}$ dimensions
using a fixed Johnson--Lindenstrauss matrix,
\begin{equation}
    \mathbf p_i=\hat{\mathbf B}_i R,
    \qquad
    R_{jl}\sim\mathcal{N}(0,1)/\sqrt{D_{\mathrm{rp}}},
    \label{eq:projection}
\end{equation}
and the logarithmic norm channels are then appended to the projection,
\begin{equation}
    \mathbf z_i
    =
    \left(\mathbf p_i,\;
    d_{i,0},\;d_{i,1},\dots,d_{i,n_{\mathrm{blocks}}}\right).
    \label{eq:feature}
\end{equation}
The norm channels do not pass through the projection, so the feature dimension
exceeds the projection dimension,
$D=D_{\mathrm{rp}}+1+n_{\mathrm{blocks}}$, corresponding to 130, 131, and 132
features for the one-, two-, and three-layer models with
$D_{\mathrm{rp}}=128$. The projection seed and the resulting matrix are stored
with the CALM parameters.

Third, the feature $\mathbf z_i$ of an atom of element $e$ is assigned to the
nearest cluster centroid of that element,
\begin{equation}
    k^{*}
    =
    \arg\min_k\lVert\mathbf z_i-\boldsymbol\mu_{e,k}\rVert^2 ,
    \label{eq:assignment}
\end{equation}
and the extrapolation grade follows from Eqs.~(\ref{eq:mahalanobis}) and
(\ref{eq:gamma}) using the mean, covariance and threshold of the assigned
cluster.

The CALM model is constructed in three streaming passes over the training
data. Each pass is evaluated on parallel workers and followed by a global
reduction, so that neither the features nor the per-atom scores of the full
training set need to be stored simultaneously.

In the first streaming pass over the dataset, each worker applies
MiniBatchKMeans separately for every element and for the candidate cluster
counts $K\in\{1,2,4,8,16\}$. The worker centroids and their populations are
then combined by weighted global KMeans for every element and candidate $K$.
The number of clusters is selected independently for each element by applying
the maximum-distance elbow criterion, the point of largest distance from
the straight line between the endpoints of the curve, to its within-cluster
sum-of-squares curve. The maximum of these elementwise choices defines $K_{\max}$, while each
element may use fewer effective clusters. Following this reduction, clusters
with an aggregated population below $D+1$ are iteratively merged into their
nearest neighbour.

In the second streaming pass over the dataset, the training environments are
reassigned to the global centroids. Each worker accumulates raw and weighted
populations and the scatter matrix for every element-cluster pair, and these
quantities are combined by addition. Because the populations estimated in the
first pass are based on worker-level centroids, the population condition is
checked again using the actual assignments. A cluster containing fewer than
$D+1$ atoms is merged into its nearest neighbour, while the centroid and
scatter matrix are combined using the parallel-axis theorem. The covariance
matrices are then formed from the reduced statistics, regularized by adding
$\varepsilon I$ with $\varepsilon=10^{-6}$, and inverted by symmetric
pseudoinversion.

In the third streaming pass over the dataset, every worker evaluates the
Mahalanobis distance $\sigma$ for its part of the training data and accumulates
raw and weighted histograms for each element-cluster pair. The histograms are
combined by addition, and the calibration threshold is obtained from the
reduced distribution as
\begin{equation}
    \theta_{e,k}
    =
    \operatorname{median}(\sigma_{e,k})
    +3\times1.4826\,
    \operatorname{MAD}(\sigma_{e,k}).
    \label{eq:calibration-threshold}
\end{equation}
The factor 1.4826 makes the MAD a consistent estimator of the standard
deviation for a normal distribution.
This choice limits the influence of the long tail of the training-set
distribution on the calibration point.
The raw histogram determines whether a cluster contains enough observations for an independent threshold. 
If it does not, its threshold is replaced by the largest reliable threshold for the same element.
Using the largest reliable threshold avoids assigning a more restrictive
boundary to a cluster with insufficient calibration data.
When source weights are applied during CALM construction, the
threshold obtained from the weighted histogram is used at inference. Both
histograms are stored with the CALM parameters and can therefore be used for later
recalibration without another model evaluation.

At inference, the nearest-centroid assignment used in
Eq.~(\ref{eq:gamma}) is treated as constant during differentiation.

The neighbourhood-averaged grade in Eq.~(\ref{eq:gamma-neighbour}) uses the
model neighbour list, including periodic images within the corresponding pair
cutoff $r_{\mathrm c,ij}$, while atom $i$ enters with unit weight. The same
operator is applied to every atomic score compared in
Table~\ref{tab:ensemble}.

For online data acquisition, newly collected environments are assigned to the
fixed centroids and used to update the cluster populations and scatter matrices.
After each collected batch, the covariance is updated according to
\begin{equation}
    C_{\mathrm{new}}
    =
    (1-\alpha)C_{\mathrm{old}}+\alpha C_{\mathrm{batch}},
\end{equation}
and its regularized inverse is recomputed.
Here $C_{\mathrm{batch}}$ is the covariance of the collected atoms
about the fixed centroid of their assigned cluster. For the aluminium
and silicon walks, we select the smallest value of
$\alpha\in\{0.05,0.1,0.2,0.4,0.7,1.0\}$ for which the maximum grade
of the collected configuration falls below the update target of 1.25.
Each trial starts from the covariance before collection. At $\alpha=1$,
the batch covariance replaces the previous covariance. If the target
is still not reached, we inflate the covariances of the clusters
represented in the configuration by at most fourfold in each of at
most three steps.
The centroids, calibration thresholds, and GRACE parameters remain fixed.

With the cluster assignment and calibration threshold held fixed, the
uncertainty-derived bias is obtained
by differentiating
\begin{equation}
    H
    =
    \sum_{i\in\mathrm{real}}\frac{\sigma_i}{\theta_i}
    =
    \sum_{i\in\mathrm{real}}\gamma_i.
\end{equation}
This derivative is omitted when only scalar values of \gam{} are required.

For comparison, D-optimality active sets are constructed with the MaxVol
algorithm~\cite{podryabinkin2017,lysogorskiy2023} on the same normalized
features. We consider one non-clustered, uncentred active set per element and one
centred active set per element-cluster pair, taking the cluster partition
from the CALM model itself, i.e. the elbow selection with $K_{\max}=8$.
An additional Supplementary control uses the same
per-cluster partition without centring. In the centred construction, the
corresponding cluster centroid is subtracted from both the training and query
features. A square
$131\times131$ active matrix is selected when the cluster contains at least 131
environments; otherwise, a rectangular matrix and its Moore--Penrose
pseudoinverse are used. With $\mathbf{x}=\boldsymbol{\varphi}$ for the
uncentred constructions and
$\mathbf{x}=\boldsymbol{\varphi}-\boldsymbol{\mu}_{e,k^{*}}$ for the centred
per-cluster construction, the D-optimality grade is
\begin{equation}
    \gamma_{\mathrm{Dopt}}
    =
    \max_j\left|\left(\mathbf{x}A^{+}\right)_j\right|.
\end{equation}

\subsection{Validation models, datasets and metrics}

The geometrical-extrapolation column of Table~\ref{tab:ensemble} is measured on
hydrostatic bond scans. For Al, Cu, Fe, Si, Ti and W, the ground-state prototype
is scaled over nearest-neighbour distances of 1.5--6.0\,\AA{} in 80 steps and
repeated until every cell vector exceeds the model cutoff. A fixed supercell is
used throughout each scan, and the maximum atomic score is evaluated for
compression ({$d_{\mathrm{nn}}<0.8\,d_{\mathrm{nn}}^{\mathrm{eq}}$}) and stretching
({$1.4\,d_{\mathrm{nn}}^{\mathrm{eq}}<d_{\mathrm{nn}}<6$~\AA}).
The CALM and D-optimality grades detect a regime when this maximum exceeds
one. The deep ensemble score is the standard deviation
of the energy per atom across the three members, and detection requires a tenfold increase of this spread over its median within 5\% of the equilibrium spacing.
A table entry reads yes only when all six prototypes are detected.
The Supplementary Information shows the complete scans of the three CALM feature variants and the sensitivity of the ensemble rule to its factor.

Method validation uses three random initializations of a two-layer GRACE model
trained on a 30,000-structure element-stratified subset of
OMat24~\cite{omat24}. The evaluation sets comprise 28{,}406 held-out
OMat24-val structures (547{,}588 atoms) and 27{,}488 SMAX
structures~\cite{smax2026} (355{,}848 atoms).
Each model uses a 6\,\AA{} cutoff, $l_{\max}=4$ and 3 in the two layers, 10
radial basis functions expanded to 42 and 32, 42 and 64 product functions, 17 density channels, maximum correlation order 4 and single precision.

Each foundation potential is paired with its own CALM model.
GRACE-3L-OMAT-large and GRACE-2L-OMAT-medium are trained on OMat24,
GRACE-2L-SMAX-OMAT-medium and GRACE-2L-SMAX-OMAT-large also include SMAX, and
GRACE-2L-SMAX-large is trained on SMAX.

Foundation-model CALM constructions use four dataset selections. The OMat24-only construction
uses 5\% of the OMat24~\cite{omat24} training data, whereas the SMAX-only
construction uses the full SMAX~\cite{smax2026} dataset. For joint
OMat24--SMAX models, the same datasets are assigned source weights of 20 and 1,
respectively. To keep the reference settings consistent with SMAX, the OMat24
subset excludes structures containing Gd, Eu, or Yb because their OMat24 and
SMAX calculations use different PAW datasets. It also excludes oxides and
fluorides containing Co, Cr, Fe, Mn, Mo, Ni, V, or W because OMat24 applies a
Hubbard $U$ to these systems while SMAX does not. 
{For the models fine-tuned on the sAlex and MPtrj data (ft-AM),} 41.25\%
of the subsampled Alexandria (sAlex)
dataset~\cite{schmidt2023alexandria,wang2023alexandria2d} and the full MPtrj
dataset~\cite{chgnet} are assigned weights of 2.4242 and 1.0. These sampling
weights enter all three passes and affect the centroids, covariances and
calibration distributions; the realized source contributions also depend on
the distribution of atomic environments and the OMat24 filter.

The fixed-threshold ROC-AUC pools all elements and labels atoms with a force
error above 1\,eV\,\AA$^{-1}$ as positive. The worst-decile protocol labels the
highest-error tenth within each element and evaluates each element separately.
Ensemble disagreement is the norm of the across-seed standard deviation of
predicted forces. Correlations and ROC-AUC values are evaluated over atoms or,
where stated, structures.

Comparative tables use the intersection of structures evaluated successfully
for every estimator and seed. This leaves 27{,}488 of 27{,}529 SMAX
structures; no OMat24-val structures are excluded.

\subsection{Controlled structural and chemical extrapolation}

The restricted Al dataset is labelled with the GRACE-2L-SMAX-OMAT-large
reference model, which supplies energies, forces and stresses.
In both controlled tests, the foundation model is used instead of DFT as the label source.
The
\emph{vacancy-free, FCC-only} dataset contains 261
structures: 11 primitive cells scanning $\pm0.5$\,\AA{} about the equilibrium
nearest-neighbour distance of 2.863\,\AA{} in steps of 0.1\,\AA{}, plus
250 perturbed $2\times2\times2$ FCC supercells (32 atoms) with the same
displacement and strain amplitudes.

The fitted potential is a one-layer GRACE model with a 6\,\AA{} cutoff and
$l_{\max}=4$, 8 radial basis functions expanded to 32, 64
product functions, 12 density channels, maximum correlation order 4, single
precision. Training uses Adam at a learning rate of 0.004 with AMSGrad and %
reduction on plateau, 600 iterations, batch size 16, a 10\% held-out test split and
Huber losses on energy, forces and stress with weights 1, 5 and 0.1, switching after
250 iterations to energy-heavy weights of 25 and 1. Its CALM model is built from
the training set with the same procedure and threshold recipe as the foundation
models.

Evaluation uses three probes. Bond scans sweep the nearest-neighbour distance from
1.0 to 6.0\,\AA{} at 160 points. Vacancies use a
$3\times3\times3$ FCC supercell of 108 atoms with one atom removed and positions
relaxed at fixed cell.
Surfaces are $3\times3$ slabs of 9 layers for the (111) and (100) facets and 11
layers for (110), with 12\,\AA{} of vacuum and periodicity in the two in-plane
directions. Layer-resolved profiles are folded by the distance to the nearest
surface.

The BCC probes use the same fixed model at its relaxed lattice constant of
3.242\,\AA{}. The vacancy contains 128 atoms in a
$4\times4\times4$ conventional cell, and the surface is a $3\times3$ (111)
cell of 15 layers.

For the chemical-extrapolation test, BCC Mo--W random solid solutions are
generated as 128-atom volume-relaxed
cells with random displacements of 0.1\,\AA{} per Cartesian component. One
relaxed volume is used at each composition. The fitted potential is a
one-layer GRACE model restricted to Mo and W, with a 6\,\AA{} cutoff,
$l_{\max}=4$, 10 radial basis functions expanded to 32, 64 product functions,
17 density channels, maximum correlation order 4 and single precision. It is
trained on 90 reference-labelled configurations with $c_{\mathrm W}\le25\%$
and evaluated on 330
configurations (42{,}240 atoms) spanning $c_{\mathrm W}=0$--$100\%$ in steps
of 10\%. GRACE-2L-SMAX-OMAT-large supplies the labels.
Its uncertainty
model is built on its training set with the normalized feature, $D=128$ and
$K=8$.

\subsection{Matbench Discovery evaluation}

The benchmark model is GRACE-3L-OMAT-large-ft-AM, which underlies the
\texttt{GRACE-3L-OAM-L} leaderboard entry~\cite{riebesell2025framework}. The
relaxed geometries and predicted formation energies are taken from the
submitted files, and \gam{} is evaluated on each relaxed geometry in one model
pass.

For the thermal-conductivity task, we evaluate $\gam$ on the relaxed
cells and recompute $\kappa_\mathrm{SRME}$ for each filtered subset by averaging
the existing per-material errors defined by P\'ota et
al.~\cite{pota2026thermal}, without further phonon calculations.

All WBM quantities describe the unique-prototype subset, 215\,488 of the
256\,961 scored structures, which is the set the leaderboard reports by
default. The discovery metrics are recomputed for each threshold-defined subset using
the published functions. For the coverage control, subsets of the same size
are drawn without replacement from that same subset, 40 times for the discovery
metrics and 200 times for conductivity.
Their mean and standard deviation define the control and its band.

\subsection{Uncertainty-biased data collection in aluminium and silicon}

In the hyperactive-learning approach~\cite{vanderoord2023}, MD is
biased along the uncertainty gradient according to
\begin{equation}
\mathbf F_i^{\mathrm{HAL}}
=\mathbf F_i^{\mathrm{phys}}+\tau_r\nabla_{\mathbf r_i}\mathcal U,
\qquad
\tau_r=\tau\,
\frac{\sum_i\lVert\mathbf F_i^{\mathrm{phys}}\rVert}
{\sum_i\lVert\nabla_{\mathbf r_i}\mathcal U\rVert},
\qquad
\mathcal U=\sum_j\gam_j,
\label{eq:hal-force}
\end{equation}
where $\mathcal U$ is the sum of the calibrated atomic grades and $\tau=0.15$.
As in the
implementation accompanying Ref.~\cite{vanderoord2023}, the two normalization
sums are exponentially averaged over ten steps. A switched short-range
ZBL repulsion, which vanishes beyond 2.4\,\AA{} in Al and 2.0\,\AA{} in Si,
is added together with its virial after the biased forces and model stress have
been formed.

The cell is biased separately by
$\boldsymbol\sigma_{\mathrm{bias}}=-\tau_s B\hat{\mathbf W}$, where
$\hat{\mathbf W}$ is the averaged unit direction of the derivative of
$\mathcal U$ with respect to cell strain. The stress scale $B$ is
0.474\,eV\,\AA$^{-3}$ for Al and 0.610\,eV\,\AA$^{-3}$ for Si. The gain
$\tau_s$ is adjusted over 2000-step windows within %
$0.005\le\tau_s\le0.10$, while the volume is confined to %
0.75--1.30\,$V_0$. At either boundary, the cell is rescaled and the isotropic
component of its momentum is reflected.

Seed datasets contain 25 decorrelated frames at 300 and 500\,K from unbiased MD
of a 108-atom FCC aluminium cell and a 128-atom diamond silicon cell. The
GRACE-1L-OMAT-large-ft-AM foundation model drives these trajectories in the
fully anisotropic Martyna--Tobias--Klein ensemble with Nos\'e--Hoover chains at
1\,bar and a 2\,fs time step, and the frames are labelled by the reference
model. A single-element GRACE-1L seed potential is fitted to each 50-frame set
using Adam for 1000 epochs and a Huber loss on energies, forces and stresses.
Its CALM model uses one cluster and 128-dimensional features. The
equilibrium volume $V_0$ is the mean over the second half of an unbiased
5000-step run of the seed potential at the collection temperature, and the
final snapshot of that run, rescaled to 0.94, 0.97, 1.03 and 1.06\,$V_0$,
seeds the four biased walks.

The walks use the same integrator, at 300\,K for aluminium and 500\,K for
silicon, for at most 100\,000 steps each after 500 equilibration steps. A
configuration is collected whenever
$\gam_{\max}$ of the live CALM model exceeds the collection threshold of 1.5, with a ten-step gap between
collections. After every collection, the CALM covariances are updated with fixed centroids
until the grade of the collected configuration falls to the update target of 1.25, and
bounded covariance inflation is applied if the update alone cannot reach this value. The GRACE
parameters remain fixed. GRACE-2L-SMAX-OMAT-large labels all collected
configurations after each walk.

From each walk we draw up to 50 collected configurations uniformly. The
unbiased collection draws 200 frames uniformly from a 20\,000-step
continuation of the foundation-model trajectories, labelled by the reference
model. Each collection is added to the 50 seed frames, and the seed potential
is upfitted from its own weights for 400 epochs.

The complete procedure is performed for two replicates of each element using
independently generated seed datasets and different random seeds throughout, so
that the replicates share only the hold-out evaluation sets. All aluminium
walks ran to the step limit, whereas several silicon walks stopped earlier or
diverged, as detailed in the Supplementary Information.
Quantities evaluated separately for
each replicate are reported as their mean and sample standard deviation,
whereas both replicates are pooled for the feature-space coverage and the
atom-resolved relation between grade and force error. For the pooled
feature-space coverage, the MD and HAL collections from both replicates are
re-evaluated in the feature space of the first replicate's seed potential
because the two seed potentials define different feature bases. For the
atom-resolved comparison, each potential is scored with its own calibrated CALM
model before pooling, so that \gam{}$=1$ defines the same calibrated boundary.

Hold-out evaluation sets generated by the reference model comprise
isothermal--isobaric trajectories between 300 and 1100\,K for aluminium and
between 300 and 2000\,K for silicon, a vacancy and a $(111)$ surface at 300 and
600\,K (500 and 900\,K for silicon), and a liquid pre-melted above the melting
point and sampled at 1100\,K (2000\,K for silicon). The vacancy and surface
entries of the figure insets contain only atoms within 6\,\AA{} of the
instantaneous defect region.

Before taking absolute values, we remove one constant per-atom energy offset per
potential and temperature, equal to its mean signed error. Coverage is compared
in the fixed atomic feature space of the seed potential. The two-dimensional
projection is fitted once on the union of the two collections, and each contour
encloses 95\% of the corresponding atomic-feature distribution.

\subsection{Foundation-model expansion}

The foundation-model expansion experiment starts from 40 chemically diverse
SMAX training structures and uses GRACE-2L-SMAX-large together with its original
CALM model. Each structure is propagated at 1000\,K while its volume is
increased linearly from $1.0V_0$ to $1.4V_0$ over at most 15{,}000 steps. The
uncertainty-force scale is $\tau=2.0$; the time step is 2\,fs, reduced to 0.5\,fs
for cells containing H, He, Li or Be. A switched ZBL repulsion provides the
short-range guard, and a configuration is rejected when its shortest bond is
below $\max(1.2\,\AA{},0.75d_{\mathrm{nn}}^0)$, where
$d_{\mathrm{nn}}^0$ is the seed nearest-neighbour distance.

Collection is triggered when at least
$\max[3,\lceil0.15N\rceil]$ atoms have a live grade $\gam>1.3$, with at least
200 MD steps between collections and at most three collected configurations per
seed. After each collection, we select the smallest value of $\alpha$
for which fewer than $\max[3,\lceil0.15N\rceil]$ atoms remain above
$\gam_i=1.3$. Each trial starts from the covariance before collection.
If no value satisfies this condition, we use the smallest value that
brings $\gam_{\max}$ below 1.25, or $\alpha=1$ if neither condition
is reached. The centroids, calibration thresholds and GRACE
parameters remain fixed.
The updated mixture is carried from one seed to the next, so a seed starts
from the coverage accumulated by the seeds before it. Two further update
protocols, restarting the mixture for every seed and never updating it, are
reported in the Supplementary Information.
For the unbiased control, the same seeds are propagated with the same
temperature, volume ramp, time-step rule and random seeds, but with $\tau=0$ and
without CALM updates. We take a control configuration from the same seed at the
nearest stored MD step available for each collected biased configuration.

The analysis in Fig.~\ref{fig:selfexp} pools 4{,}810
atoms from 78 biased configurations and 4{,}206 atoms from 69
unbiased configurations. The grade is the neighbourhood-averaged
\(\gnbr_i\) of Eq.~(\ref{eq:gamma-neighbour}).

\subsection{DFT single-point protocol}

Reference calculations for the foundation-model expansion are static,
spin-polarized single points evaluated with VASP 5.4.4~\cite{kresse1996vasp}
following the SMAX
protocol~\cite{smax2026}. We use the PBE exchange-correlation
functional~\cite{perdew1996pbe}, the VASP-recommended PBE PAW potentials selected
for each element in the SMAX protocol~\cite{kresse1999paw}, a plane-wave cutoff
of 500\,eV, a $\Gamma$-centred mesh with a k-spacing of 0.125\,\AA$^{-1}$,
Gaussian smearing of 0.1\,eV, $\texttt{EDIFF}=10^{-6}$\,eV,
\texttt{PREC=Accurate}, and no Hubbard $U$. Initial magnetic moments follow the
SMAX protocol.

\subsection{LAMMPS overhead measurements}

Overhead is measured on W BCC systems of 2{,}000 and
16{,}000 atoms on a
single A100 GPU, for the TensorFlow \texttt{grace} and Kokkos
\texttt{grace/\{1,2,3\}l/kk} pair styles, as the ratio of MD loop times
with and without the per-atom \gam{}
computation requested through the LAMMPS fix-pair mechanism
(\texttt{fix uq all pair $N$ grace\ldots\ gamma 1}).
Each timing runs two warm-up steps followed by five timed NVE steps on a single
MPI rank, and every configuration is repeated three times. Requesting the
\texttt{uncertainty\_force} field from the same fix selects the model signature
that also evaluates the gradients of $\sigma$ with respect to the atomic
positions. The models are the GRACE-1L-OMAT-large-ft-E,
GRACE-2L-OMAT-medium-ft-E, GRACE-2L-OMAT-large-ft-E and GRACE-3L-OMAT-large
checkpoints.

\section*{Data availability}
The CALM models are distributed together with the GRACE foundation potentials
that provide \gam{}. The potentials are available at
\url{https://huggingface.co/AMS-ICAMS-RUB/grace-foundation-models} and documented
at \url{https://gracemaker.readthedocs.io/en/latest/gracemaker/foundation/}.

\section*{Code availability}
The CALM extrapolation-grade implementation is available in GRACEmaker version
0.6.0 at \url{https://github.com/ICAMS/grace-tensorpotential/releases/tag/0.6.0}.
Documentation and usage examples are available at
\url{https://gracemaker.readthedocs.io/en/latest/gracemaker/uq/}.

\bibliographystyle{unsrt}
\bibliography{references}

\section*{Acknowledgements}
 High-performance computing (HPC) resources were provided by the Paderborn Center for Parallel Computing (PC$^2$) and the Elysium HPC cluster at Ruhr-Universit\"at Bochum.
The authors acknowledge the use of Claude (Anthropic), ChatGPT (OpenAI) and Gemini (Google) for assistance with literature search, editing and as a formulation aid. The authors take full responsibility for all scientific arguments and final content. 

\section*{Funding}
We acknowledge funding from the Deutsche Forschungsgemeinschaft (DFG, German Research Foundation) through SFB 1394 (project number 409476157).

\section*{Author contributions}
Conceptualisation and Project Administration: All authors. 
Y.L. and A.B. developed the software and parameterized the models.
Writing - original draft: Y.L. 
Writing-review and editing: All authors.
Resources and funding acquisition: Y.L., R.D.

\section*{Competing interests}
Y.L., A.B. and R.D. hold equity interests in ACEworks GmbH, which provides
commercial services related to the development and application of ACE and
GRACE interatomic potentials. The authors declare no other competing interests.

\clearpage
\thispagestyle{empty}
\begin{center}
{\Large\bfseries Supplemental information for\par}
\vspace{0.7em}
{\LARGE\bfseries A latent-space extrapolation grade built into graph atomic
cluster expansion foundation potentials\par}
\vspace{1.4em}
{\large Yury Lysogorskiy$^{1,*}$, Anton Bochkarev$^{1}$, Ralf Drautz$^{1}$\par}
\vspace{0.8em}
{\normalsize $^{1}$ICAMS, Ruhr-Universit\"at Bochum, Bochum, Germany\par}
\vspace{0.4em}
{\normalsize $^{*}$Corresponding author:
\href{mailto:yury.lysogorskiy@rub.de}{yury.lysogorskiy@rub.de}\par}
\vspace{0.4em}
{\normalsize \today\par}
\end{center}
\setcounter{figure}{0}
\setcounter{table}{0}
\renewcommand{\thefigure}{S\arabic{figure}}
\renewcommand{\thetable}{S\arabic{table}}
\setcounter{equation}{0}
\renewcommand{\theequation}{S\arabic{equation}}
\newcounter{sinote}

\newcommand{\sisubsec}[2]{%
  \FloatBarrier
  \refstepcounter{sinote}%
  \phantomsection
  \subsection*{SI\arabic{sinote}. #2}%
  \label{si:#1}%
}
\newenvironment{sitoc}{\small\begin{list}{}{%
  \setlength{\leftmargin}{1.2em}\setlength{\labelwidth}{0pt}%
  \setlength{\topsep}{0pt}\setlength{\partopsep}{0pt}%
  \setlength{\itemsep}{0.12em}\setlength{\parsep}{0pt}}}{\end{list}}
\newcommand{\sitocline}[2]{\item[]\hyperref[#1]{#2}\dotfill\pageref{#1}}
\newcommand{\sitocfloat}[3]{%
  \item[]\hyperref[#1]{#2~\ref*{#1}. #3}\dotfill\pageref{#1}}

\vspace{1.6em}
{\noindent\large\bfseries Contents\par}
\vspace{0.6em}

{\noindent\bfseries Supplementary Notes}
\vspace{0.2em}
\begin{sitoc}
\sitocline{si:collide}{SI1. Construction of colliding environment pairs}
\sitocline{si:stretch}{SI2. Mechanism of stretching blindness in the linear feature}
\sitocline{si:featspace}{SI3. Feature space, the choice of $D$ and $K$, and calibration diagnostics}
\sitocline{si:pooled}{SI4. Neighbourhood-averaged extrapolation grade}
\sitocline{si:mow}{SI5. Geometric and initialization controls for Mo--W extrapolation}
\sitocline{si:founddiag}{SI6. Additional diagnostics for foundation models}
\sitocline{si:halprotocol}{SI7. Details of uncertainty-biased collection and analysis}
\sitocline{si:sihal}{SI8. Uncertainty-biased data collection in silicon}
\sitocline{si:dftctrl}{SI9. DFT and collection controls for foundation-model expansion}
\sitocline{si:cost}{SI10. Runtime overhead of the built-in extrapolation grade}
\sitocline{si:estim}{SI11. Density- and leverage-based estimators on identical data}
\end{sitoc}

\vspace{0.9em}
{\noindent\bfseries Supplementary Figures}
\vspace{0.3em}
\begin{sitoc}
\sitocfloat{fig:enn-ref}{Figure}{Uncertainty along bond scans of the reference model, for three latent-feature variants}
\sitocfloat{fig:pca2d}{Figure}{Clusters in the uncertainty feature space}
\sitocfloat{fig:grid}{Figure}{Contamination curves}
\sitocfloat{fig:grid05}{Figure}{Contamination of the confident set at a fixed force-error threshold}
\sitocfloat{fig:aucsweep}{Figure}{Detection of large force errors}
\sitocfloat{fig:atomic-reliability}{Figure}{Reliability of the calibrated threshold for the atomic grade}
\sitocfloat{fig:atomic-dfhist}{Figure}{Training-data dependence of foundation-model uncertainty, for the atomic grade}
\sitocfloat{fig:pooled-algeom}{Figure}{Spatial resolution of the neighbourhood-averaged grade}
\sitocfloat{fig:selfexp-bare}{Figure}{Atomic-grade control for foundation-model data collection}
\sitocfloat{fig:enn-all}{Figure}{Energy and per-atom uncertainty along bond-length scans, element by element}
\sitocfloat{fig:sihal}{Figure}{Uncertainty-biased data collection at a matched labelling budget in silicon}
\sitocfloat{fig:dopt}{Figure}{Density- versus leverage-based uncertainty on identical data}
\end{sitoc}

\vspace{0.9em}
{\noindent\bfseries Supplementary Tables}
\vspace{0.3em}
\begin{sitoc}
\sitocfloat{tab:quantiles}{Table}{Measured force-error distribution in each \gam{} bin}
\sitocfloat{tab:auc-comparison}{Table}{Large-force-error detection across feature spaces and estimators}
\sitocfloat{tab:pooling-applications}{Table}{Effect of neighbourhood averaging in the application tests}
\sitocfloat{tab:chem}{Table}{Mo--W initialization ablation}
\sitocfloat{tab:selfexpdft}{Table}{Collection and DFT completion for the foundation-model expansion}
\sitocfloat{tab:cost}{Table}{Runtime overhead of per-atom \gam{} in LAMMPS}
\sitocfloat{tab:baselines}{Table}{Distance-metric ablation in the normalized feature space}
\sitocfloat{tab:llpr}{Table}{Last-layer prediction rigidity on the hidden-layer feature}
\end{sitoc}
\clearpage

\sisubsec{collide}{Construction of colliding environment pairs}

The non-uniqueness of the hidden-layer mapping (main text) was demonstrated by
explicit counterexamples. Pairs of environments with coinciding central-atom
hidden features were constructed by minimizing $\lVert\mathbf z_B-\mathbf
z_A\rVert^{2}$ over the atomic positions of a second cluster of $M=50$
neighbours (Adam followed by L-BFGS-B, with minimum-distance penalties),
starting either from a perturbed copy of the first environment or from an
unrelated random cluster. From unrelated initializations the optimization
converges to relative feature differences of order $10^{-9}$ while the two
geometries remain about 5.9\,\AA{} apart in RMSD after optimal rigid-body
superposition, i.e.\ with translational and rotational degrees of freedom
compensated, and their sorted interatomic-distance sets differ as well, which
excludes a hidden symmetry relating the pair. For the control at $M=6$
neighbours, the kinematic dimension $3M-3=15$ is below the 17 density
channels, so this dimension count does not require distinct geometries
to share hidden features. In the tested matching runs, the residual
stalls six orders of magnitude higher. A complementary null-space walk,
with rotations projected out and Gauss--Newton projection back onto the feature
level set, moves the geometry by up to about 1\,\AA{} RMSD at a feature drift
below $10^{-9}$. For the tested configurations, the joint Jacobian
of all per-atom features has rank $3N-6$. Attempts to match the features
of every atom from a distant initialization either fail or return to
the original geometry.

\sisubsec{stretch}{Mechanism of stretching blindness in the linear feature}

\begin{figure}[!ht]
\centering
\includegraphics[width=0.92\textwidth]{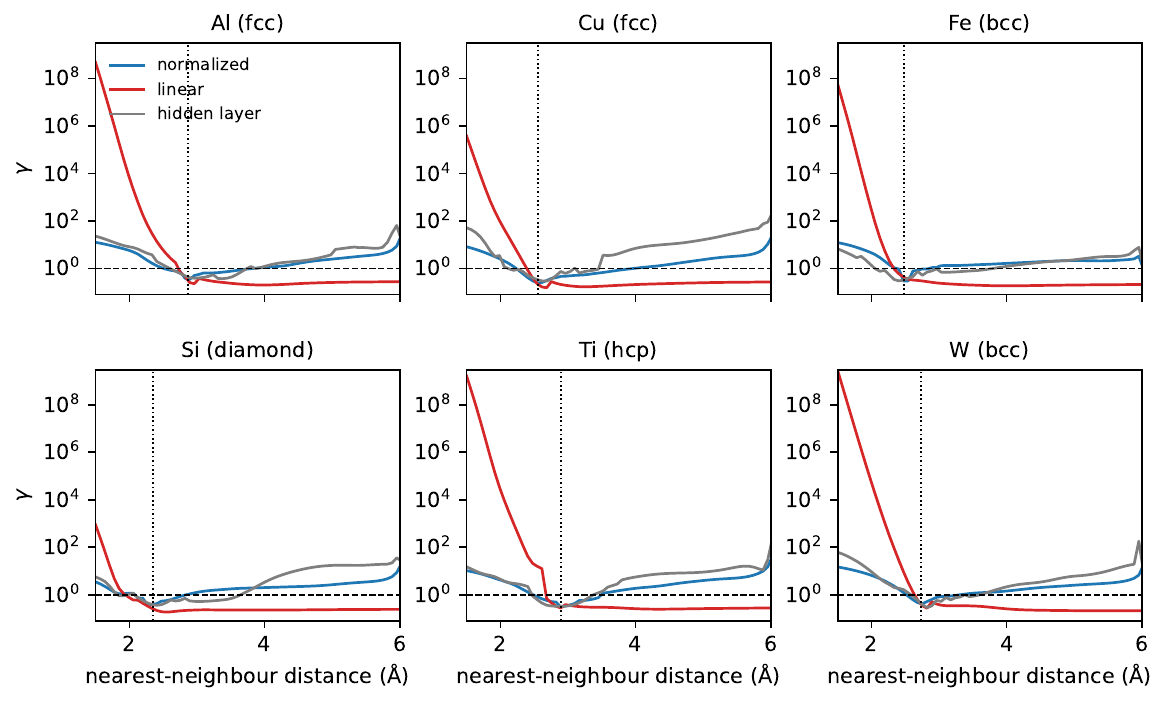}
\caption{\textbf{Uncertainty along bond scans of the reference model, for three
latent-feature variants.} Per-atom \gam{}, on a logarithmic scale, along
hydrostatic nearest-neighbour-distance scans of the six elemental ground-state
prototypes, scored with the controlled reference fit. The three variants share
the same energy model and differ only in the latent feature the mixture is built
on. The dotted vertical line marks the equilibrium spacing and the dashed
horizontal line the calibration threshold $\gam=1$. Each panel ends at the
6\,\AA{} cutoff.}
\label{fig:enn-ref}
\end{figure}

Figure~\ref{fig:enn-ref} shows that the linearly projected basis fails to
identify stretching because its feature vector approaches the origin as the
last neighbour leaves the cutoff. Since the training feature norms span several
orders of magnitude, the dominant covariance direction describes variations in
feature magnitude and assigns only a small Mahalanobis penalty to displacement
towards the origin. The symmetry of the Mahalanobis distance may therefore
allow the large-norm compression tail to broaden the model also towards
vanishing norms. This interpretation is consistent with the endpoint
decomposition, although it has not been isolated by refitting after removal of
the compression tail. In comparison, the normalized feature preserves the
stretching response through its logarithmic norm channels.
For the deep ensemble, the smallest increase over the median energy spread
within 5\% of the equilibrium spacing is 49-fold under stretching and 99-fold
under compression. Thus, all six prototypes remain detected for any multiplier
below 49, including the tenfold criterion used in main-text Table~1.

\sisubsec{featspace}{Feature space, the choice of $D$ and $K$, and calibration diagnostics}

\begin{figure}[!ht]
\centering
\makebox[\textwidth][c]{\includegraphics[width=1.10\textwidth]{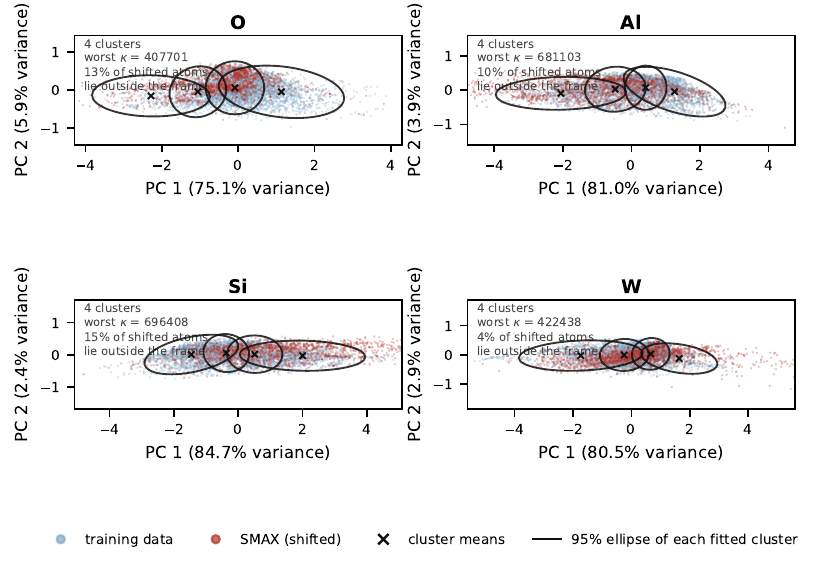}}
\caption{\textbf{Clusters in the uncertainty feature space.} Two-dimensional
PCA projections of the normalized feature for GRACE-2L-OMAT-medium-ft-E. Blue
points are OMat24 training atoms, red points are SMAX atoms, black crosses are
cluster means, and ellipses show projected 95\% Gaussian contours. PCA is fitted
separately for each element and the panels use equal aspect, so their axes are
not directly comparable. Each panel gives the number of clusters, the largest
covariance condition number, and the fraction of SMAX atoms outside the plotted
training-data range.}
\label{fig:pca2d}
\end{figure}

\begin{figure}[!ht]
\centering
\includegraphics[width=\textwidth]{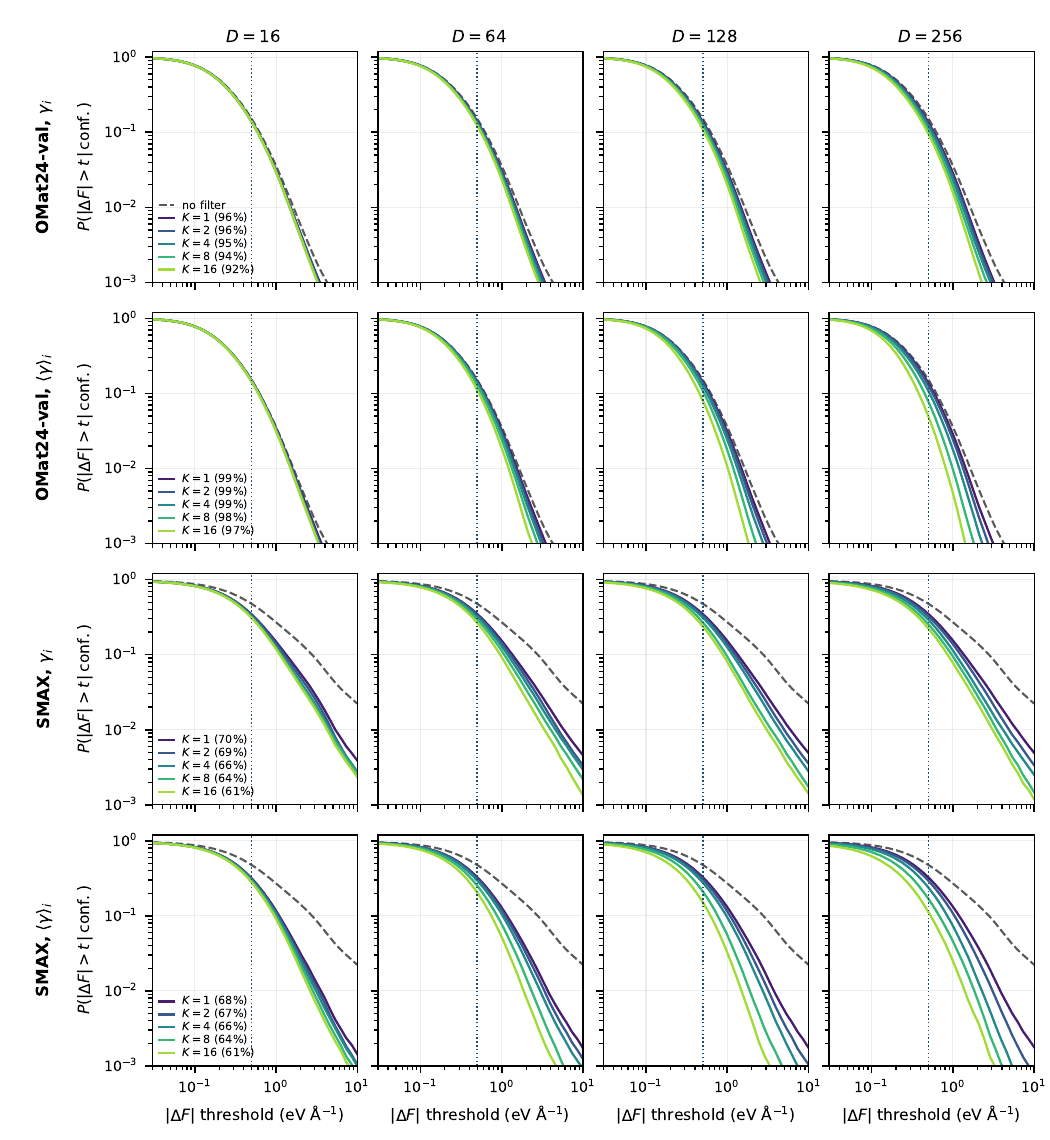}
\caption{\textbf{Contamination curves.} Probability that an atom classified as
reliable has a force error above a variable threshold, for the controlled GRACE-2L
reference model trained on the 30,000-structure element-stratified OMat24 subset.
Columns give the projection dimension $D$ and solid curves the cluster count $K$, with
legend percentages giving the fraction of atoms kept by the filter. Rows pair each
evaluation set with each grade, the atomic \(\gam_i\leq1\) and the
neighbourhood-averaged \(\gnbr_i\leq1\), in the order OMat24-val then SMAX. The
dashed curve labelled \emph{no filter} is the force-error exceedance probability over
all atoms before filtering, and the dotted vertical line marks the
$0.5\,\text{eV\,\AA}^{-1}$ threshold tabulated in Fig.~\ref{fig:grid05}.}
\label{fig:grid}
\end{figure}

\begin{figure}[!ht]
\centering
\includegraphics[width=\textwidth]{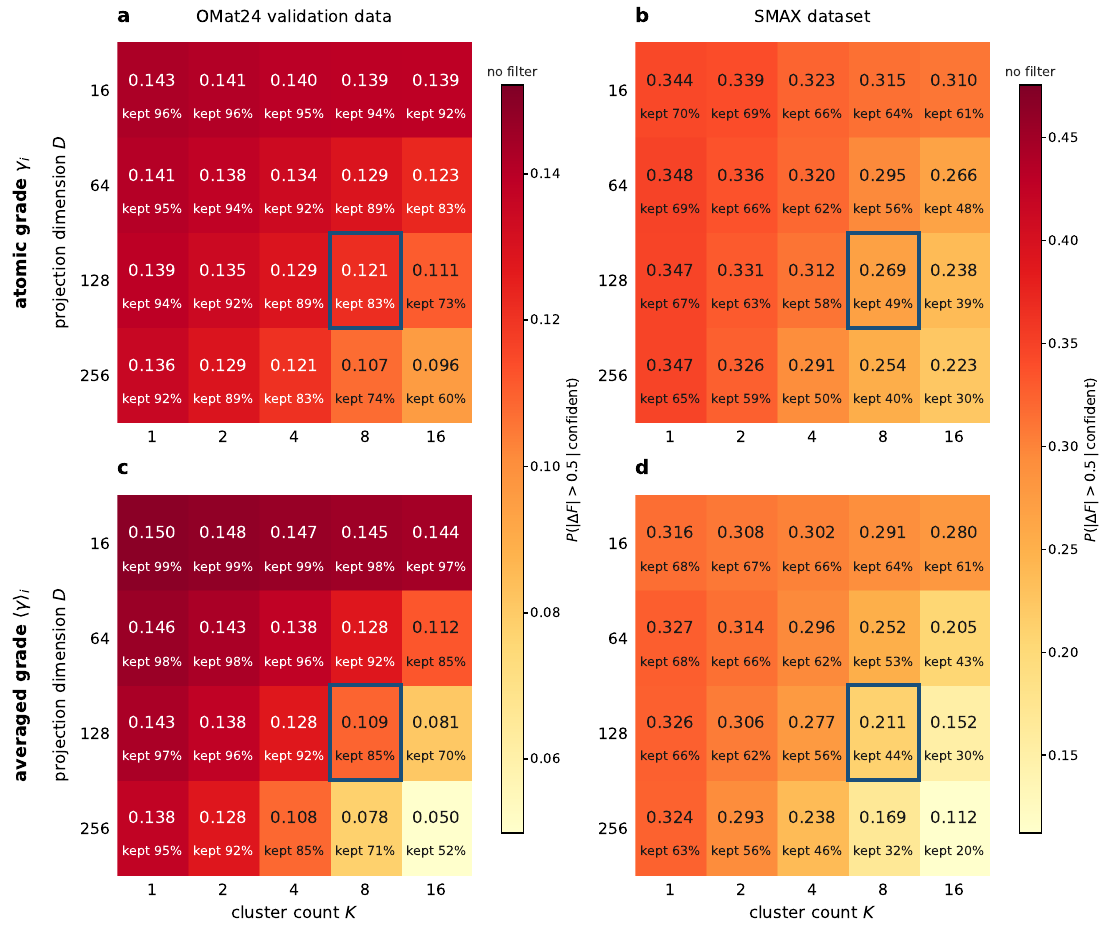}
\caption{\textbf{Contamination of the confident set at a fixed force-error
threshold.} Fraction of the atoms called reliable whose force error against DFT
nevertheless exceeds $0.5\,\text{eV\,\AA}^{-1}$, over the same grid as
Fig.~\ref{fig:grid} and for the same controlled reference fit, on
\textbf{a}~held-out OMat24-val data and \textbf{b}~the SMAX dataset for the atomic
grade $\gam_i\leq1$, and on \textbf{c}~OMat24-val and \textbf{d}~SMAX for the
neighbourhood-averaged grade $\gnbr_i\leq1$. The
large number in each cell is that probability, and the small number below it is
retention, the fraction of atoms the filter keeps. The colour scale is shared down
each column, so the two grades are directly comparable. The top of each colour bar
is the unfiltered probability over all atoms before filtering,
0.15 on OMat24-val data and
0.48 on SMAX. The blue box marks $D=128$ and $K=8$. The
foundation models keep $D=128$ and select the cluster count per element by the
elbow criterion, which gives $K_{\max}=8$ with most elements using four
clusters.}
\label{fig:grid05}
\end{figure}

Figures~\ref{fig:pca2d}--\ref{fig:grid05} show the clustered feature space of a
foundation model and the effect of the projection dimension $D$ and the cluster
count $K$ on the contamination of the confident set.

For the atomic grade, contamination decreases with $K$ at every $D$,
and with $D$ for $K\geq2$, while the
fraction of atoms kept decreases as well. The lowest-contamination cells keep
only 60\% and 30\% of the atoms. At
$D=128$ and $K=8$, the probability of a force error above
$0.5\,\text{eV\,\AA}^{-1}$ among the confident atoms decreases to
0.12 on OMat24-val data and 0.27 on
SMAX, while 83\% and 49\% of the atoms are
kept.

The following diagnostics use the atomic grade $\gam_i$. The corresponding
neighbourhood-averaged analysis is reported in the main text.
The reliability analysis in the main text reports bin medians because a small
number of atoms can dominate the mean in the sparsely populated high-\gam{}
bins. The quoted
Spearman correlations are instead computed over individual atoms, avoiding the
inflation that would result from correlating binned averages. The uncertainty
bands widen at high \gam{} as the score distribution thins from approximately
one hundred thousand atoms to a few hundred atoms per bin.

\begin{figure}[!ht]
\centering
\includegraphics[width=\textwidth]{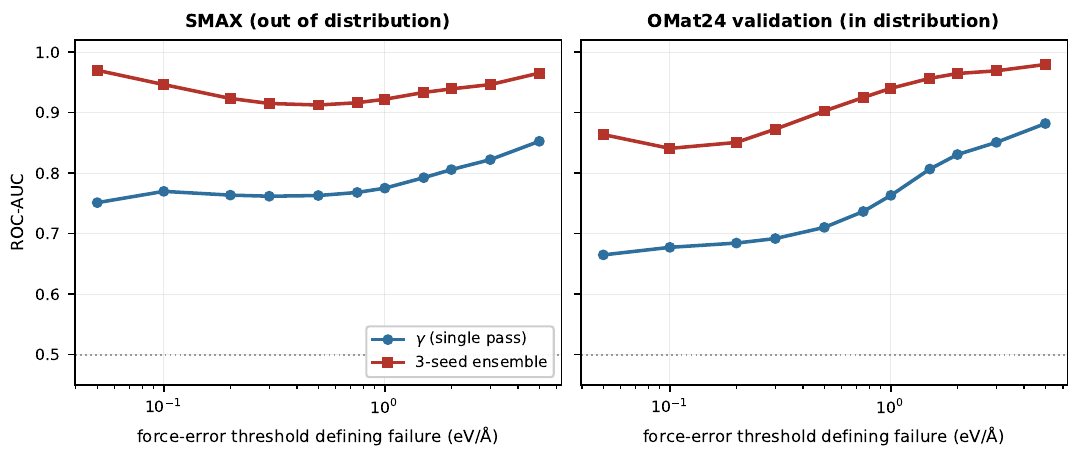}
\caption{\textbf{Detection of large force errors.} ROC-AUC for identifying atoms
whose force error exceeds a threshold, against that threshold, for the
controlled reference model. The thresholds span 0.05 to
5.0\,eV\,\AA$^{-1}$. The dotted line marks chance.}
\label{fig:aucsweep}
\end{figure}

The mean force error increases across every populated \gam{} bin on both
evaluation sets. It rises from 0.083 to
6.30\,eV\,\AA$^{-1}$ on OMat24-val and from
0.038 to 9.43\,eV\,\AA$^{-1}$ over
11 populated bins on SMAX. The atom-level association is
stronger on SMAX ($\rho=0.539$) than on OMat24-val
($\rho=0.376$), where the score spans a narrower range. On SMAX,
the median rises from 0.031 to
1.1\,eV\,\AA$^{-1}$ up to $\gam\approx3$, over bins containing
93.6\% of the atoms, and then saturates. The continued increase of
the mean and upper quantiles therefore comes from a growing high-error tail.
Figure~\ref{fig:aucsweep} and Table~\ref{tab:quantiles} provide the corresponding
threshold sweep and conditional error quantiles.
Across the reported thresholds, the ROC-AUC of \gam{} increases overall from
0.751 to 0.853 on SMAX and from
0.665 to 0.882 on OMat24-val,
with a shallow non-monotonicity on SMAX. The ensemble has a higher ROC-AUC at
every threshold and is approximately constant on SMAX. Thus, the difference
decreases by approximately half for the largest errors but does not close.
Table~\ref{tab:auc-comparison} reports the same detection metric for every
feature space and estimator of main-text Table~1, for both the atomic and the
neighbourhood-averaged grade.

\begin{table}[!ht]
\centering
\caption{\textbf{Measured force-error distribution in each \gam{} bin.}
Conditional force-error distributions observed at each \gam{} for the
controlled reference model. We report percentiles of the per-atom force error
in eV\,\AA$^{-1}$, the share of atoms in each bin, and the share exceeding
1\,eV\,\AA$^{-1}$. The median changes modestly in distribution, while the 99th
percentile increases by two orders of magnitude. Thus, increasing \gam{}
identifies a growing probability and magnitude of large force errors.}
\label{tab:quantiles}
\begin{tabular}{llrrrrr}
\toprule
set & $\gam{}$ bin & atoms (\%) & median & p90 & p99 & $>1$\,eV\,\AA$^{-1}$ (\%) \\
\midrule
SMAX & 0--0.5 & 13.3 & 0.133 & 0.531 & 1.35 & 2.5 \\
 & 0.5--0.75 & 24.1 & 0.307 & 1.038 & 3.38 & 10.7 \\
 & 0.75--1 & 17.4 & 0.424 & 1.574 & 6.62 & 19.5 \\
 & 1--1.5 & 16.4 & 0.585 & 2.776 & 10.83 & 32.5 \\
 & 1.5--2 & 9.4 & 1.086 & 4.190 & 14.27 & 52.7 \\
 & 2--5 & 16.8 & 1.052 & 5.673 & 34.55 & 51.8 \\
 & $>5$ & 2.6 & 3.161 & 31.281 & 118.49 & 65.4 \\
\midrule
OMat24-val & 0--0.5 & 33.5 & 0.136 & 0.373 & 0.87 & 0.6 \\
 & 0.5--0.75 & 36.7 & 0.216 & 0.605 & 1.35 & 2.7 \\
 & 0.75--1 & 16.7 & 0.269 & 0.791 & 1.98 & 6.0 \\
 & 1--1.5 & 9.8 & 0.316 & 0.953 & 2.57 & 9.1 \\
 & 1.5--2 & 2.2 & 0.382 & 1.173 & 3.44 & 13.5 \\
 & 2--5 & 1.1 & 0.429 & 1.397 & 8.29 & 17.2 \\
 & $>5$ & 0.1 & 0.377 & 1.559 & 154.13 & 17.9 \\
\bottomrule
\end{tabular}
\end{table}

\begin{table}[!ht]
\centering
\caption{\textbf{Large-force-error detection across feature spaces and
estimators.} ROC-AUC for detecting atoms with force errors above
$1\,\mathrm{eV}\,\text{\AA}^{-1}$ on OMat24-val and SMAX. The methods and
evaluation atoms are identical to those in main-text Table~1, and each entry is
reported as $\gam_i/\gnbr_i$. The bold feature name marks the selected CALM
construction.}
\label{tab:auc-comparison}
\begin{tabular}{@{}clcc@{}}
\toprule
 & Feature space & OMat24-val & SMAX \\
 & & ROC-AUC & ROC-AUC \\
\midrule
\multicolumn{2}{l}{Deep ensemble ($\times3$)}
  & 0.940/0.935
  & 0.922/0.900 \\
\midrule
\multirow{3}{*}{\rotatebox[origin=c]{90}{CALM}}
  & \textbf{Normalized feature}
  & 0.763/0.843
  & 0.775/0.814 \\
  & Linear projected basis
  & 0.754/0.833
  & 0.819/0.835 \\
  & Hidden layer
  & 0.680/0.780
  & 0.772/0.830 \\
\midrule
\multirow{3}{*}{\rotatebox[origin=c]{90}{D-opt}}
  & \shortstack[l]{Normalized, centred\\per cluster}
  & 0.718/0.818
  & 0.771/0.816 \\
  & \shortstack[l]{Normalized, uncentred\\per element}
  & 0.671/0.753
  & 0.746/0.798 \\
  & \shortstack[l]{Hidden layer, uncentred\\per element}
  & 0.609/0.681
  & 0.741/0.798 \\
\bottomrule
\end{tabular}
\end{table}

\sisubsec{pooled}{Neighbourhood-averaged extrapolation grade}

Main-text Table~1 compares the atomic grade \(\gam_i\) with the
neighbourhood-averaged grade \(\gnbr_i\) defined in the main text. The averaging
improves the force-error indication of every single-pass scheme by between 0.05
and 0.16 in rank correlation. The deep ensemble is the one exception and
decreases slightly, since its atomic force spread is already localized to the
force error being compared. The ordering of the methods remains unchanged.

The precise averaging weights have little influence on this result. Replacing
the cutoff weights by a uniform average over the same neighbours changes the
rank correlation by at most 0.005, while varying the radius from the first
coordination shell to the model cutoff changes it by at most 0.003. The
geometrical test of main-text Table~1 is also unchanged because all sites in
each ideal crystal are equivalent.

The uncentred per-cluster D-optimality control omitted from main-text Table~1
follows the same trend. Its correlation and ROC-AUC change from 0.242 and 0.703
to 0.402 and 0.811 on OMat24-val, and from 0.485 and 0.756 to 0.581
and 0.802 on SMAX. Thus, the improvement does not depend on the centred
D-optimality construction used in the main comparison.

\begin{figure}[!ht]
\centering
\includegraphics[width=\textwidth]{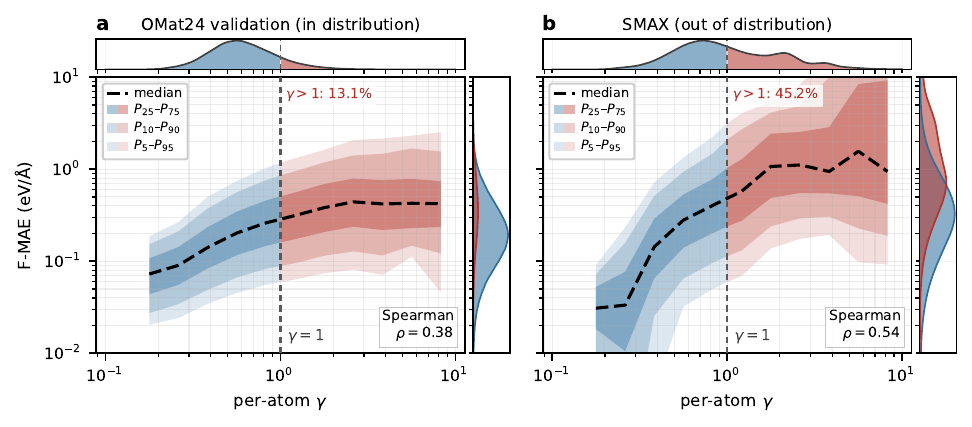}
\caption{\textbf{Reliability of the calibrated threshold for the atomic grade.}
As main-text Fig.~2, with the abscissa given by the atomic grade \gam{} instead of
its neighbourhood average, on \textbf{a}~held-out OMat24-val data and
\textbf{b}~SMAX. The dashed line marks \gam{}$=1$, blue and red the two sides of it,
the black dashed curve the bin median, and the nested bands the $P_{25}$--$P_{75}$,
$P_{10}$--$P_{90}$ and $P_{5}$--$P_{95}$ intervals. Each panel gives the fraction of its
atoms above the boundary, evaluated on its own cache rather than on the shared comparison
set of Table~1, and the
probability of a force error above $0.5\,\text{eV\,\AA}^{-1}$ rises from
0.13 to 0.32 across the boundary on OMat24-val
and from 0.30 to 0.69 on
SMAX, against the averaged values quoted in the main text.}
\label{fig:atomic-reliability}
\end{figure}

\begin{figure}[!ht]
\centering
\includegraphics[width=0.86\textwidth]{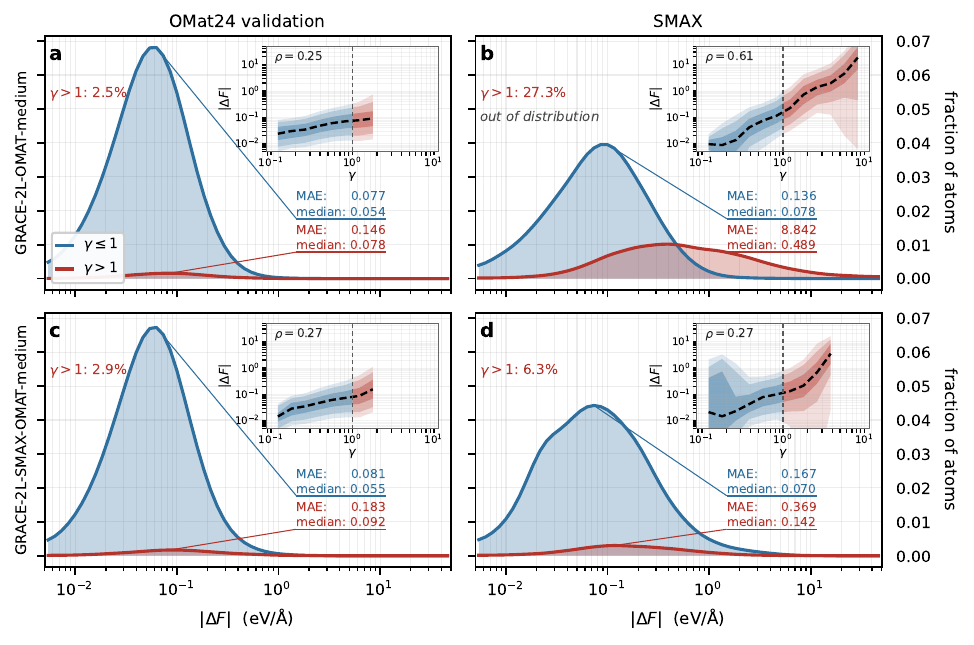}
\caption{\textbf{Training-data dependence of foundation-model uncertainty, for the
atomic grade.} As main-text Fig.~5, splitting each panel at \gam{}$=1$ instead of at
the neighbourhood-averaged boundary. Rows correspond to GRACE-2L-OMAT-medium and
GRACE-2L-SMAX-OMAT-medium, columns to OMat24-val and SMAX, and the insets
give the force error against \gam{} with the per-atom Spearman correlation. On SMAX
the OMat24-only model flags 27.3\% of atoms against
6.3\% for the model that has seen SMAX. Both populations are
visible here, which is why this version is kept alongside the averaged one.}
\label{fig:atomic-dfhist}
\end{figure}

\begin{figure}[!ht]
\centering
\includegraphics[width=\textwidth]{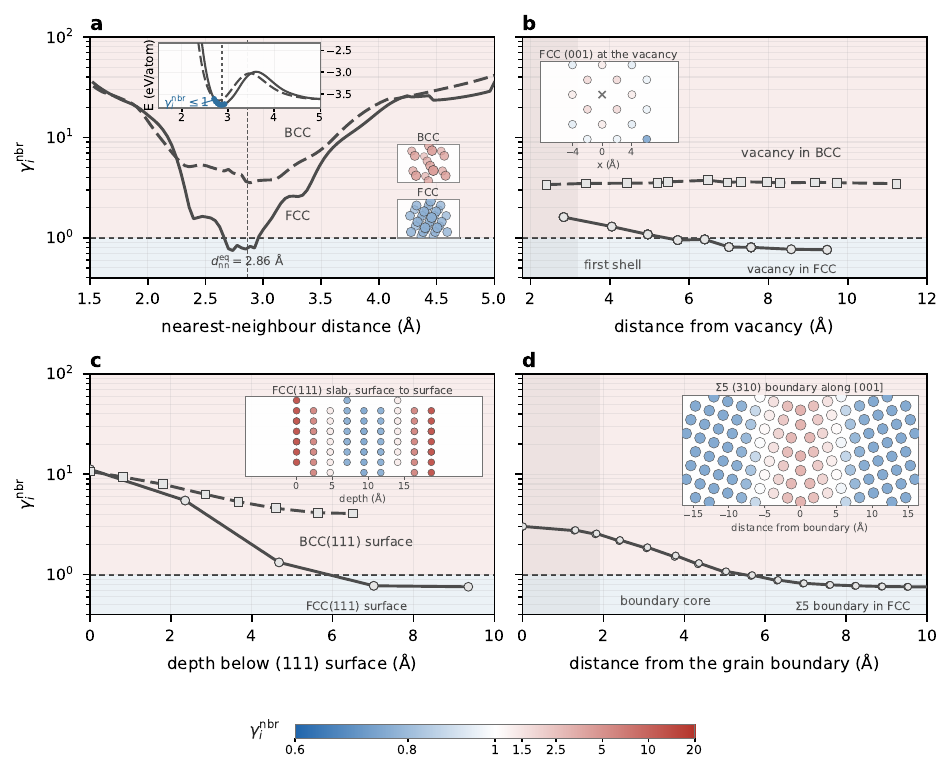}
\caption{\textbf{Spatial resolution of the neighbourhood-averaged grade.} As
main-text Fig.~3, with \(\gnbr_i\) in place of the atomic grade, and shaded by the
same boundary.
\textbf{a}~Bond-length scans, unchanged by the averaging because every site
of an ideal crystal is equivalent. \textbf{b}~Monovacancy, \textbf{c}~free surface
and \textbf{d}~grain boundary, where the averaging lowers the peak values and
widens the flagged region.}
\label{fig:pooled-algeom}
\end{figure}

\begin{figure}[!ht]
\centering
\includegraphics[width=\textwidth]{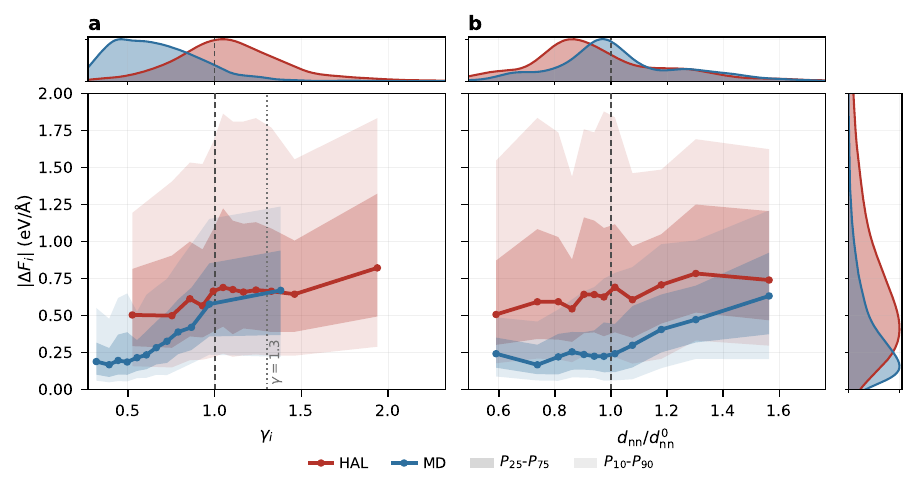}
\caption{\textbf{Atomic-grade control for uncertainty-biased collection on a
foundation model.} The configurations and DFT errors are identical to those in
main-text Fig.~8, but the extrapolation grade is the atomic value \(\gam_i\).
\textbf{a}~Atomic force error against \(\gam_i\). \textbf{b}~The same error
against the normalized nearest-neighbour distance.}
\label{fig:selfexp-bare}
\end{figure}

Neighbourhood averaging lowers contamination at $D=128$ and $K=8$ on
both evaluation sets, although this improvement does not hold over the
complete grid (Figs.~\ref{fig:grid} and \ref{fig:grid05}). It also spreads
a localized defect signal over the neighbouring atoms
(Fig.~\ref{fig:pooled-algeom}). The atomic-grade counterparts of the three
main-text figures that use the averaged grade are given as
Figs.~\ref{fig:atomic-reliability}, \ref{fig:atomic-dfhist} and
\ref{fig:selfexp-bare}, while Table~\ref{tab:pooling-applications} summarizes the
remaining application controls without repeating the complete main-text figures.

For OMat24-val and SMAX, neighbourhood averaging increases the
rank correlations from 0.376 to 0.497 and from
0.539 to 0.617, respectively. The fractions above
the calibrated boundary change from 13.1\% to
10.2\% and from 45.1\% to
48.5\%. Both grades are evaluated at a cutoff of one.
These comparisons use the adopted CALM model, whose cluster count is
selected per element.

For the atomic grade at $D=128$ and $K=8$ in the grid with fixed $K$,
the contamination of the confident set is
0.12 on OMat24-val and 0.27 on SMAX, with
83\% and 49\% of the atoms kept.
This grid requests the same $K$ for every element, so its kept fractions
are not complementary to the flagged fractions of the adopted model.
Around an FCC vacancy, averaging moves
the crossing of unity from 4.5 to approximately 6\,\AA{} and broadens the
flagged region. In the foundation-model collection test, the two error curves
obtained with the atomic grade approach one common relation after
neighbourhood averaging.

\begin{table}[!ht]
\centering
\caption{\textbf{Effect of neighbourhood averaging in the application tests.}
The entries compare the atomic and neighbourhood-averaged grades on identical
configurations.}
\label{tab:pooling-applications}
\begin{tabular}{>{\raggedright\arraybackslash}p{0.24\textwidth}
                >{\raggedright\arraybackslash}p{0.29\textwidth}
                >{\raggedright\arraybackslash}p{0.39\textwidth}}
\toprule
analysis & measured change & interpretation \\
\midrule
Mo--W chemical extrapolation
  & atomic-force correlation: 0.464 to 0.452
  & nearly unchanged because the atomic grade already follows the local W concentration \\
\hline  
Foundation-model domains
  & correlations increase in all four model--dataset pairs
  & the contrast between training domains increases, while only 0.4--0.5\% of OMat24-val atoms remain above the boundary \\
\hline  
WBM reliability filtering
  & flagged structures: 20.1\% to 2.8\%; error above the boundary: 23.2 to 30.8\,meV/atom
  & unchanged filtering conclusion for cells with a median of eight atoms \\
\hline  
Al data acquisition
  & atomic-force correlation: 0.74 to 0.83 for the MD %
    potential and 0.65 to 0.78 for the HAL potential
  & the selected configurations are unchanged because collection used \(\gam_i\) \\
\bottomrule
\end{tabular}
\end{table}

\sisubsec{mow}{Geometric and initialization controls for Mo--W extrapolation}

Each Mo--W composition is relaxed separately. The volume per atom increases
from 15.858\,\AA$^3$ for pure Mo to 16.217\,\AA$^3$ for pure W, a
change of 2.26\% over the complete composition range. This
corresponds to a 0.75\% change in the cubic lattice parameter
(3.1654 to 3.1891\,\AA), or 0.075\% for each 10\% step in
W content. In the elemental bond-length scans, \gam{} changes by only a few
percent over strains of this magnitude and rises appreciably only for bond
changes of approximately ten percent. The severalfold increase across the
Mo--W series is therefore dominated by chemical rather than geometrical
novelty.

Table~\ref{tab:chem} compares the two initializations. Both show similar
detection quality, while their accuracy and score baselines differ. The
fine-tuned model is more accurate at pure W but preserves a broader feature
distribution from the foundation model. Therefore, its local-coordination score
already exceeds one at pure Mo, whereas the model trained from random
initialization starts well below one and spans a wider range. Fine-tuning thus
improves extrapolative accuracy while giving a less distinct boundary for this
axis-specific test.

\begin{table}[!ht]
\centering
\caption{\textbf{Mo--W initialization ablation.} Both models are trained
on the same $c_{\mathrm W}\le25\%$ data and differ only in initialization.}
\label{tab:chem}
\begin{tabular}{lcc}
\toprule
metric & random init & fine-tuned \\
\midrule
force MAE @ 0\% / 100\% W (eV\,\AA$^{-1}$) & 0.045 / 0.099 & 0.036 / 0.071 \\
degradation $0\!\to\!100\%$ W & $2.2\times$ & $2.0\times$ \\
mean \gam{} $0\!\to\!100\%$ W & $0.61\to1.96$ & $0.61\to1.84$ \\
\gam{} @ local W $=0$ / $8$ & 0.75 / 1.83 & 1.08 / 1.72 \\
Spearman / ROC-AUC & 0.464 / 0.803 & 0.438 / 0.793 \\
\bottomrule
\end{tabular}
\end{table}

\sisubsec{founddiag}{Additional diagnostics for foundation models}

For GRACE-3L-OMAT-large on SMAX (374{,}701 atoms), the ROC-AUC
is 0.935 for detecting per-atom force errors above
1\,eV\,\AA$^{-1}$ and 0.874 at a 0.5\,eV\,\AA$^{-1}$
threshold. The median \gam{} is 0.680,
17.8\% of atoms lie above $\gam=1$, and the Spearman
correlation between \gam{} and the force error is 0.487. On
OMat24-val (547{,}588 atoms), the median is
0.556, 2.7\% of atoms lie above the
same boundary, and the correlation is 0.165.

\begin{figure}[!ht]
\centering
\includegraphics[width=\textwidth]{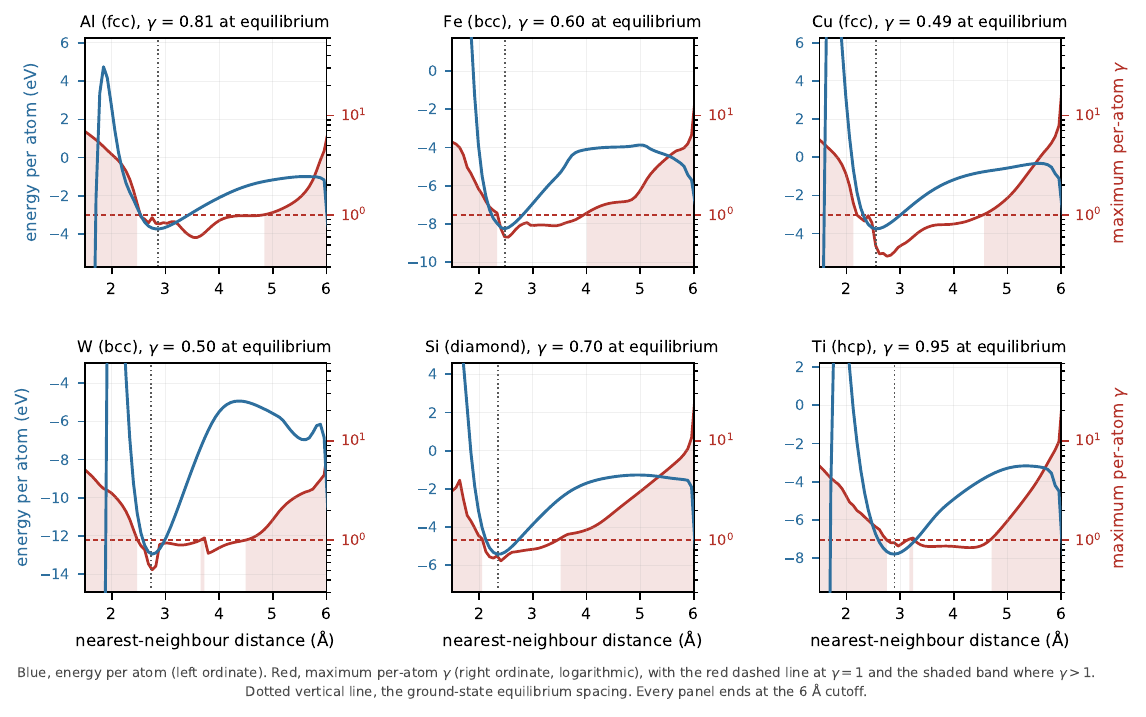}
\caption{\textbf{Energy and per-atom uncertainty along bond-length scans, element by
element.} Energy per atom (blue, left ordinate) and maximum per-atom \gam{}
(red, logarithmic right ordinate) for GRACE-3L-OMAT-large along hydrostatic
bond-length scans of six elemental prototypes. The dotted line marks the
ground-state spacing, the dashed line marks \gam{}$=1$, and the shaded region
has \gam{}$>1$. Each panel ends at the 6\,\AA{} cutoff. The extrapolation grade remains
below one around equilibrium and rises under compression and stretching.}
\label{fig:enn-all}
\end{figure}

Figure~\ref{fig:enn-all} applies the geometric test of the main text to the
same model. Along hydrostatic bond scans of six elemental prototypes, \gam{}
remains below one around the ground-state spacing and rises above it under both
compression and stretching. The foundation model therefore passes the same
geometric test as the reference fits of main-text Table~1.

The two GRACE-2L foundation models compared in the main text differ in both
their trained weights and their support sets. These effects cannot be separated
by fitting the extrapolation grade to each model's own training dataset, so their
different extrapolation rates establish an empirical association with training
coverage rather than a causal decomposition. The two SMAX panels also contain
different atom sets. GRACE-2L-OMAT-medium covers 83 of the
88 elements present and skips structures containing At,
Fr, Po, Ra, or Rn. It is therefore evaluated on 374{,}701 of the
390{,}994 atoms used for GRACE-2L-SMAX-OMAT-medium, corresponding
to 95.8\% coverage.

For the thermal-conductivity benchmark, removing the higher-\gam{} materials
leaves fewer materials in the average, which by itself may lower
$\kappa_\mathrm{SRME}$. We therefore compare the filtered set at
\gam{}$\leq0.5$ with 200 size-matched random subsets of the same
103 materials, which give
$\kappa_\mathrm{SRME}=0.1210\pm0.0062$ (mean and standard
deviation) against 0.1041 for the filtered set. Only
1 of the random subsets reaches an equal or lower error, i.e.\ a
one-sided empirical $p=0.010$, so the improvement reported in the main
text is not a coverage effect.
The set mean of $\kappa_\mathrm{SRME}$ over the 103 materials is
0.1211, which reproduces the published value for this entry, and only
2 of them exceed $\gam=1$.

\sisubsec{halprotocol}{Details of uncertainty-biased collection and analysis}

A stress normalized relative to the physical stress does not drive the cell
efficiently, because the elastic stress increases with the deformation caused
by the bias. We therefore use the fixed-magnitude stress defined in the main
Methods. Its gain $\tau_s$ is adjusted from the volume drift over a 2000-step
window: it is multiplied by 1.3 when the drift is below the target pace and
halved when the drift is too fast or the volume approaches within 5\% of a
boundary. At a volume boundary, the cell is rescaled onto the boundary and the
isotropic component of the cell momentum is reflected, while the deviatoric
component is preserved.

For the aluminium and silicon walks, the CALM update strength is
selected by a line search with a maximum-grade target of 1.25.
If the covariance update does not reach this target, bounded
covariance inflation is applied as described in the main Methods.
A frozen copy of the seed CALM is evaluated alongside the live model
but does not affect the dynamics. All eight aluminium walks ran the full
100\,000 steps, as did two silicon walks in the first replicate and one in the
second. The other two silicon walks of the first replicate were stopped
manually after about 4000 steps, when each already held more than 250
configurations, five times the 50 drawn from each walk. In the second
replicate, a walk is stopped automatically once it holds at least 50 collected
configurations and has touched a volume boundary three times within 2000
steps, which ended two silicon walks after 3768 and 4447 steps. The remaining
silicon walk of this replicate diverged under the bias after 5277 steps, and
the 365 configurations collected before that point remain in the pool from
which the 50 are drawn.

As in the controlled tests, the reference model takes the place of DFT as the
label source, and its own extrapolation grade on the collected configurations,
$\gam_{\max}\le0.61$ in aluminium and $\gam_{\max}\le1.03$ in silicon, is
recorded as an observation that the labelled configurations remain within its
training support. This grade is monitored but enters neither the collection
criterion nor the fitted potentials.

For the defect evaluation, the surface region is measured from the outermost
atoms of each non-periodic slab. The vacancy centre is located separately in
each frame as the point that maximizes the distance to the $k$-th nearest atom,
with $k=1$ for the FCC cell and $k=5$ for the diamond cell, where the nearest
distance alone does not distinguish the vacant site from a tetrahedral
interstitial void. Nearly every atom carrying the largest grade lies within
the resulting 6\,\AA{} region, so this restriction changes the local force
errors but has little effect on the maximum grades.

For the energy comparisons, the mean signed per-atom error is removed
independently for each potential and temperature before absolute errors are
formed. Coverage is evaluated atom by atom in the fixed feature space of the
seed potential. The two-dimensional projection is fitted to the union of the
two collections, and all held-out configurations are projected into the same
plane. Each 95\% region is obtained from a two-dimensional histogram of the
projected features, smoothed with a Gaussian kernel on a grid padded beyond the
data range.

\sisubsec{sihal}{Uncertainty-biased data collection in silicon}

\begin{figure}[!ht]
\centering
\includegraphics[width=\textwidth]{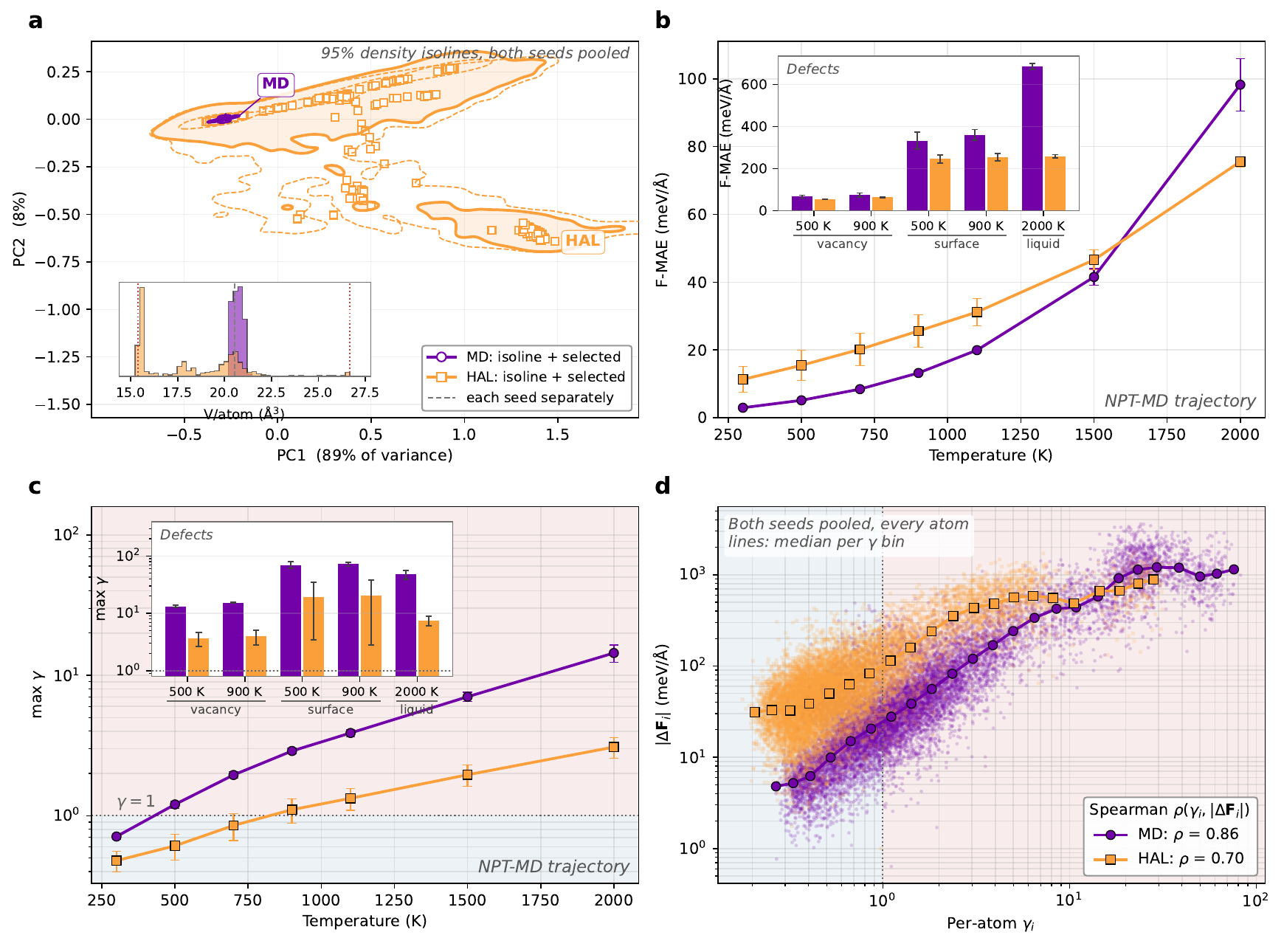}
\caption{\textbf{Uncertainty-biased data collection at a matched labelling
budget in silicon.} For each replicate, the 50 reference-labelled seed
configurations are extended by 200 configurations from either unbiased
foundation-model MD at 500\,K (MD, purple circles) or four uncertainty-biased
walks with the collection rule $\gam_{\max}>1.5$ (HAL, orange squares). A
GRACE-1L potential is fitted to each dataset. The two replicates use
independently generated seed datasets and different random seeds throughout.
Curves in \textbf{b} and \textbf{c} give the mean of the two replicates, and
the bars give their sample standard deviation.
\textbf{a}~The two collections in the first two principal components of the
seed potential's features, with the replicates pooled. Solid isolines enclose
95\% of the pooled atomic distributions, and dashed isolines show the two
replicates separately. Symbols mark the selected-configuration centroids, and
the inset gives the volume-per-atom histograms.
\textbf{b}~Force MAE against the reference model on independent reference
trajectories between 300 and 2000\,K, and, in the inset, on the held-out
vacancy, $(111)$-surface and liquid sets, with the vacancy and surface values
restricted to atoms within 6\,\AA{} of the defect.
\textbf{c}~Maximum per-atom grade on the same configurations, scored with each
potential's own CALM model, over regions shaded for $\gam\leq1$ (blue) and
$\gam>1$ (red).
\textbf{d}~Per-atom force error against the extrapolation grade \gam{} for all
evaluation sets combined and both replicates pooled, shaded by the same
boundary. Lines with symbols give the median per grade bin.}
\label{fig:sihal}
\end{figure}

The silicon counterpart of the aluminium experiment in main-text Fig.~7 uses
the same protocol. The seed dataset contains 50 reference-labelled frames of a
128-atom diamond cell from unbiased foundation-model MD, and 200 additional
configurations are collected either from a continuation of these trajectories
or from four uncertainty-biased walks at 500\,K. As for aluminium, we repeat
the complete workflow with an independently generated seed dataset and new
random seeds throughout. Fig.~\ref{fig:sihal} combines the two replicates using
the same averaging and pooling as main-text Fig.~7.

The high-temperature improvement is more pronounced than in aluminium. The
force-error crossover lies between 1500 and 2000\,K,
and on the molten cell at 2000\,K the error decreases from
687 to 258\,meV\,\AA$^{-1}$ for the
potential fitted to the uncertainty-guided collection. At 900\,K, the surface
error within 6\,\AA{} of the faces decreases from 361 to
254\,meV\,\AA$^{-1}$, although neither collection
contains a surface. The extrapolation grades follow the same trend. The
uncertainty-guided potential crosses $\gam=1$ between 700
and 900\,K, against 300 to
500\,K for the unbiased potential. The Spearman correlations
between the per-atom grade and force error, averaged over the two replicates as
for aluminium, are 0.87 and 0.77 for the MD and
uncertainty-guided potentials, respectively (Fig.~\ref{fig:sihal}c,d).

However, the broader coverage gives a larger near-equilibrium accuracy loss
than in aluminium. In both replicates, the biased walks migrate to the
compression boundary because their direction is determined by the volume
derivative of the seed potential's summed grade at $V_0$. The resulting
collection is therefore dominated by strongly compressed cells. Below about
1500\,K, the potential fitted to this collection is less accurate than the
potential fitted to unbiased MD, and near room temperature it is also less
accurate than the 50-frame seed potential. The two replicates differ more
strongly in silicon, which may reflect the larger cell distortions reached by
the second set of walks. Overall, the broader training distribution improves
the high-temperature and surface accuracy but reduces the near-equilibrium
accuracy.

\sisubsec{dftctrl}{DFT and collection controls for foundation-model expansion}

For the foundation-model expansion, the DFT protocol was checked directly on
37 of the 40 SMAX seed structures. The force-component MAE relative to the
stored SMAX labels is $0.0031\,\text{eV\,\AA}^{-1}$, less than 1\% of the
model error on the unbiased control. All grades in this analysis are recomputed
with the original CALM model instead of the mixture updated during collection.

The collection was run under three protocols for the CALM covariance
matrices. In the first the mixture is updated after every collection and carried
from one seed to the next, in the second it is restarted for every seed, and in
the third it is never updated. We ran the first two protocols for all 40 seeds
and the third for a subset of 12 seeds. Each protocol collected at most three
configurations per seed.
Table~\ref{tab:selfexpdft} gives the configurations submitted for DFT and the
number labelled. The main text uses the carried-mixture collection and the
available thermal controls matched to this protocol. The other two update
protocols enter this section as controls.
Across all three protocols, 187 of 225 uncertainty-biased configurations
completed. The configurations that did not complete are the most expanded cells
of the collection, so the labelled set understates the grades and the force
errors that the biased search reached.

\begin{table}[!ht]
\centering
\caption{\textbf{Collection and DFT completion for the foundation-model
expansion.} Configurations submitted for DFT single-point calculations and the number
for which the calculation completed, per protocol for the CALM covariance
matrices. The unbiased-control row combines the distinct thermal frames matched
by seed and nearest stored MD step to labelled configurations from all three
update protocols. A thermal frame shared by more than one protocol enters once.}
\label{tab:selfexpdft}
\begin{tabular}{lccc}
\toprule
CALM update protocol & seeds run & submitted & labelled \\
\midrule
carried across seeds   & 40 & 94  & 78 \\
restarted for each seed & 40 & 96  & 79 \\
never updated          & 12 & 35  & 30 \\
\midrule
unbiased control       & 37 & 156 & 143 \\
\bottomrule
\end{tabular}
\end{table}

The configuration-level comparison here uses 71 pairs for which the
magnetic initialization exactly follows the SMAX convention. Each pair has
the same seed, temperature schedule and volume schedule, with the unbiased
frame taken at the nearest stored MD step. An earlier magnetic
initialization was corrected to the SMAX table for part of the set, and on the
24 cells labelled under both, the correction changes the force MAE of a cell by
at most $0.03\,\text{eV\,\AA}^{-1}$ and the energy by at most 1\,meV per atom.
These pairs are drawn from all three update protocols, whereas the main text
uses the carried mixture alone. Across these matched
configuration pairs, the median configuration-level force MAE is
$0.680\,\text{eV\,\AA}^{-1}$ for uncertainty-biased sampling and
$0.390\,\text{eV\,\AA}^{-1}$ for the unbiased control. Since the force-loss term
used in training is defined from absolute errors, the biased configurations
make larger contributions to the training objective.

The carried and per-seed CALM updates draw their collections from 35
and 37 productive seeds, that is, seeds for which at least one configuration
met the selection rule, respectively.
Their median original-mixture grades and
DFT force errors are nearly identical, whereas the minimum-bond filter rejects
170 configurations for the carried mixture and 1247 when the mixture is
restarted for every seed. Without online updates, the median atom-index overlap
between consecutive collections increases from about 0.19 to 0.49. In the
fixed-volume control without the short-range safeguards, the median shortest
bond at collection contracts to 1.21\,\AA{}, compared with a median seed
nearest-neighbour distance of 1.34\,\AA{}. These controls support carrying the
mixture between seeds, keeping the online update, and opening the tensile route
with a volume ramp.

\sisubsec{cost}{Runtime overhead of the built-in extrapolation grade}

Table~\ref{tab:cost} gives the measured cost of evaluating \gam{} inside a
LAMMPS trajectory, for both backends, four foundation models and two system
sizes.

\begin{table}[!ht]
\centering
\caption{\textbf{Runtime overhead of per-atom \gam{} in LAMMPS.} Percentage
change in MD loop time relative to evaluation of energies and forces alone,
for BCC tungsten on a single A100 GPU (80\,GB). Entries give the mean and
sample standard deviation over three runs, with \gam{} evaluated at every
step. TensorFlow is timed for \gam{} alone and together with
$\partial\sigma/\partial\mathbf r$, while Kokkos is timed for \gam{} alone.
OOM denotes a run that exceeded the available GPU memory.}
\label{tab:cost}
\begin{tabular}{llccc}
\toprule
 & & \multicolumn{3}{c}{Change in MD loop time (\%)} \\
\cmidrule(lr){3-5}
 & & \multicolumn{2}{c}{TensorFlow} & Kokkos \\
\cmidrule(lr){3-4}\cmidrule(lr){5-5}
model & atoms & \gam{} & \gam{} $+\,\partial\sigma/\partial\mathbf{r}$ & \gam{} \\
\midrule
\multirow{2}{*}{1L-large} & 2{,}000 & $5.7\,{\pm}\,0.7$ & $32.3\,{\pm}\,1.5$ & $4.0\,{\pm}\,0.2$ \\
 & 16{,}000 & $3.0\,{\pm}\,0.4$ & $33.7\,{\pm}\,0.4$ & $3.3\,{\pm}\,0.3$ \\
\addlinespace
\multirow{2}{*}{2L-medium} & 2{,}000 & $4.6\,{\pm}\,0.6$ & $34.3\,{\pm}\,0.8$ & $1.7\,{\pm}\,0.1$ \\
 & 16{,}000 & $1.7\,{\pm}\,0.4$ & $33.9\,{\pm}\,0.5$ & $1.5\,{\pm}\,0.6$ \\
\addlinespace
\multirow{2}{*}{2L-large} & 2{,}000 & $2.6\,{\pm}\,0.3$ & $29.6\,{\pm}\,0.5$ & $1.2\,{\pm}\,0.5$ \\
 & 16{,}000 & $1.4\,{\pm}\,0.8$ & $30.6\,{\pm}\,0.6$ & $1.1\,{\pm}\,0.8$ \\
\addlinespace
\multirow{2}{*}{3L-large} & 2{,}000 & $0.3\,{\pm}\,0.3$ & $76.6\,{\pm}\,0.4$ & $0.7\,{\pm}\,0.1$ \\
 & 16{,}000 & $-0.2\,{\pm}\,0.3$ & $\text{OOM}$ & $0.7\,{\pm}\,0.4$ \\
\bottomrule
\end{tabular}
\end{table}

\sisubsec{estim}{Density- and leverage-based estimators on identical data}

\begin{figure}[!ht]
\centering
\includegraphics[width=0.85\textwidth]{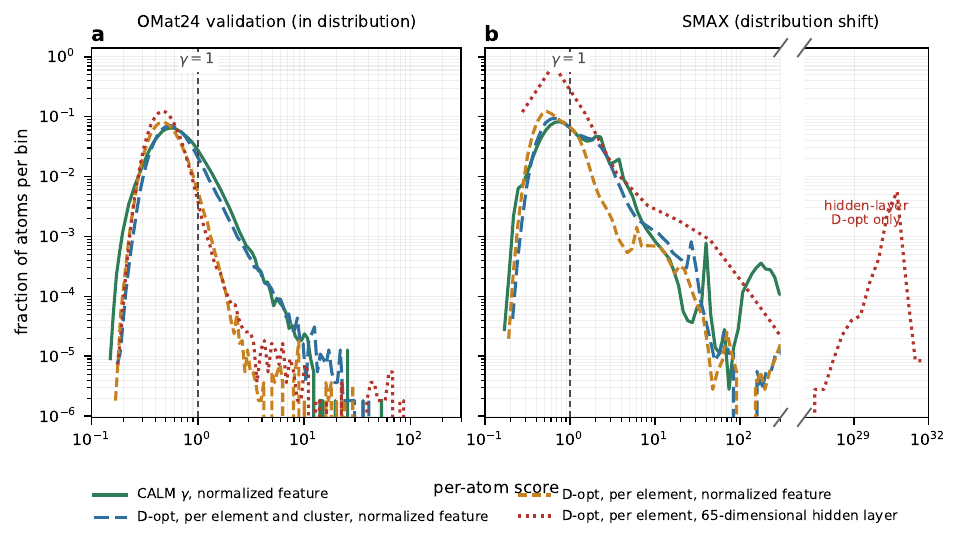}
\caption{\textbf{Density- versus leverage-based uncertainty on identical
data.} Per-atom score distributions for the CALM \gam{} and three
D-optimality variants on \textbf{a}~OMat24-val data and
\textbf{b}~SMAX, for the controlled reference model. The axis break in
\textbf{b} separates the per-element hidden-feature D-optimality scores from
the other distributions.}
\label{fig:dopt}
\end{figure}

Figure~\ref{fig:dopt} and Table~\ref{tab:baselines} compare the CALM \gam{}
with D-optimality leverage scores and with simpler distance
metrics, all built on the same normalized feature and scored on identical atoms.

On SMAX, per-element D-optimality built on the 65-dimensional hidden-layer
feature reaches $9.6\times10^{31}$. In comparison, per-cluster
D-optimality on the normalized feature and the CALM grade remain bounded, with
maxima of $1.6\times10^{4}$ and $9.9\times10^{3}$, respectively. The
separation spans approximately twenty-eight orders of magnitude, precluding a
common operating threshold for the hidden-feature score. All estimators in
Fig.~\ref{fig:dopt} are evaluated on the same 356{,}080 SMAX atoms,
which is a slightly larger set than the 355{,}848-atom intersection in
main-text Table~1 and explains the different maxima.

\begin{table}[!ht]
\centering
\caption{\textbf{Distance-metric ablation in the normalized feature space.}
Per-atom Spearman correlation with the force error and ROC-AUC for detecting
per-atom force errors above 1\,eV\,\AA$^{-1}$, for the controlled reference model,
on the same evaluation sets and with the same force-error definition as
main-text Table~1. CALM-K8, the per-cluster Mahalanobis distance with eight
clusters requested for every element; CALM-K1, the same construction with one full-covariance
Gaussian per element; and Euclidean, the distance to the nearest training
centroid. All three use the same normalized features and evaluation atoms
(351{,}200 SMAX and 516{,}681 OMat24-val atoms). Bold
marks the best value in each column. The values differ from main-text
Table~1 because this ablation requests eight clusters for every element, while
the model of main-text Table~1 selects the count per element by the elbow
criterion, and because the two are evaluated on different atom sets.}
\label{tab:baselines}
\begin{tabular}{lcccc}
\toprule
 & \multicolumn{2}{c}{$\rho(\gam,\,|\Delta\mathbf F|)$}
 & \multicolumn{2}{c}{AUC @1\,eV\,\AA$^{-1}$} \\
\cmidrule(lr){2-3}\cmidrule(lr){4-5}
method & SMAX & OMat24-val & SMAX & OMat24-val \\
\midrule
CALM-K8 & \textbf{0.556} & \textbf{0.368}
        & 0.786 & \textbf{0.766} \\
CALM-K1 & 0.492 & 0.335
        & 0.753 & 0.734 \\
euclid  & 0.536 & 0.171
        & \textbf{0.822} & 0.718 \\
\bottomrule
\end{tabular}
\end{table}

The Euclidean baseline does not use the per-cluster threshold calibration, so
Table~\ref{tab:baselines} compares the constructions as used. The fraction of
atoms above the $1\,\text{eV\,\AA}^{-1}$ error threshold is
27.2\% on SMAX and 3.7\% on
OMat24-val, and the two ROC-AUC columns therefore describe different
detection problems. Normalizing the Euclidean distance by the elementwise RMS
training radius does not improve the correlations
($\rho=0.503$ and $0.131$), so the
raw distance is reported.

Last-layer prediction rigidity (LLPR)~\cite{bigi2024,chong2025} uses the hidden
layer of the energy readout rather than the normalized feature. Because the
atomic energy is linear in this layer, the structure energy is linear in the sum
of its atomic feature vectors $\mathbf h_i$, and the training data define an
uncentred Gram matrix over structures $s$,
\begin{equation}
    G
    =
    \sum_{s}\mathbf x_s\mathbf x_s^{\mathsf T},
    \qquad
    \mathbf x_s=\frac{1}{N_s}\sum_{i\in s}\mathbf h_i ,
    \label{eq:si-llpr-gram}
\end{equation}
where $N_s$ is the number of atoms in structure $s$. The per-atom normalization
in Eq.~(\ref{eq:si-llpr-gram}) reproduces the per-atom energy normalization used
in the training loss. The atomic uncertainty is then
\begin{equation}
    u_i
    =
    \sqrt{\mathbf h_i^{\mathsf T}\left(G+\lambda I\right)^{-1}\mathbf h_i},
    \qquad
    \lambda=\eta\,\frac{\operatorname{tr}G}{D},
    \label{eq:si-llpr}
\end{equation}
with a single regularization parameter $\eta$ that is varied over $10^{-12}$ to
$10^{-2}$. Different from CALM, this construction is uncentred and uses one
global matrix per model rather than a partition over elements and clusters,
since the readout weights are shared across species.

Table~\ref{tab:llpr} reports this model on the same 65-dimensional readout hidden layer that carries the CALM
and D-optimality rows of main-text Table~1. On SMAX the correlation is weaker
than for either of those two constructions, and on OMat24-val the ROC-AUC is
close to the value expected from random ranking. Neighbourhood averaging
improves both columns, as it does for every atomic score in this comparison,
but does not change the ordering. Note that the raw structure sum ranks
somewhat better than the per-atom design row, even though the latter is the one
that reproduces the energy normalization used during training, so both are
reported.

For the GRACE architecture used in this comparison, the readout hidden
layer is a deterministic function of the $D_{\rho}=17$ density bottleneck
(main text). The LLPR, CALM and D-optimality scores constructed on this
layer therefore cannot distinguish environments with identical hidden
features. This shared representational limit motivates constructing the
GRACE extrapolation grade on the product basis before the density contraction.

\begin{table}[!ht]
\centering
\caption{\textbf{Last-layer prediction rigidity on the hidden-layer feature.}
Per-atom Spearman correlation with the force error, its neighbourhood-averaged
counterpart, and ROC-AUC for detecting per-atom force errors above
1\,eV\,\AA$^{-1}$, for the controlled reference model. LLPR denotes the
per-atom design row of Eq.~(\ref{eq:si-llpr-gram}); LLPR-sum denotes the same
construction with the raw structure sum in place of the per-atom mean. Both use
the 65-dimensional readout hidden layer, the same evaluation sets and the same
force-error definition as main-text Table~1, and are scored on the same
355{,}848 SMAX and 547{,}588 OMat24-val atoms.}
\label{tab:llpr}
\begin{tabular}{lcccccc}
\toprule
 & \multicolumn{2}{c}{$\rho(u,\,|\Delta\mathbf F|)$}
 & \multicolumn{2}{c}{$\rho(u_{\mathrm{nbr}},\,|\Delta\mathbf F|)$}
 & \multicolumn{2}{c}{AUC @1\,eV\,\AA$^{-1}$} \\
\cmidrule(lr){2-3}\cmidrule(lr){4-5}\cmidrule(lr){6-7}
method & SMAX & OMat24-val & SMAX & OMat24-val & SMAX & OMat24-val \\
\midrule
LLPR     & $0.111$ & $-0.052$
         & $0.189$ & $0.022$
         & $0.595$ & $0.514$ \\
LLPR-sum & $0.178$ & $-0.002$
         & $0.256$ & $0.069$
         & $0.621$ & $0.538$ \\
\bottomrule
\end{tabular}
\end{table}

Two properties of this construction limit how far the values in
Table~\ref{tab:llpr} can be pushed. First, the regularization parameter cannot
be selected in a meaningful way here. Every metric varies monotonically with
$\eta$ over the range examined, so that the weakest regularization on the grid
is always preferred, and the correlation on SMAX falls from
$0.111$ at $\eta=10^{-12}$ to $-0.050$ at
$\eta=10^{-2}$. The values quoted above are therefore taken at the
grid endpoint rather than at an interior optimum.

Second, the training Gram matrix is poorly conditioned. Its condition number is
$4.5\times 10^{13}$, and only 43 of its 65 directions carry more
than $10^{-6}$ of the largest eigenvalue. At the selected regularization the
shift $\lambda$ is $4.8\times 10^{-9}$, which is smaller than the smallest
eigenvalue of $6.0\times 10^{-9}$. The ranking may therefore depend on directions
that the underlying single-precision model evaluation does not resolve, which
is a further reason to read Table~\ref{tab:llpr} as an upper estimate of what
this construction achieves on the hidden-layer feature.

\end{document}